\documentclass[%
 reprint,
 amsmath,amssymb,
 aps,
]{revtex4-1}
\usepackage[utf8]{inputenc}
\usepackage[english]{babel}

\usepackage{graphicx, float}
  \graphicspath{{pictures/}}
\usepackage{dcolumn}
\usepackage{bm}

\usepackage{subfiles}  
\usepackage{xcolor} 

\begin{document}
 \title{Excitation of a bound state in the continuum via spontaneous symmetry breaking}

\author{Alexander Chukhrov}
  \affiliation{Department of Physics and Engineering, ITMO University, Saint Petersburg 197101, Russia}%

\author{Sergey Krasikov}%
  \email{s.krasikov@metalab.ifmo.ru}
  \affiliation{Department of Physics and Engineering, ITMO University, Saint Petersburg 197101, Russia}%

\author{Alexey Yulin}
  \affiliation{Department of Physics and Engineering, ITMO University, Saint Petersburg 197101, Russia}%

\author{Andrey Bogdanov}
   \email{a.bogdanov@metalab.ifmo.ru}
  \affiliation{Department of Physics and Engineering, ITMO University, Saint Petersburg 197101, Russia}%

\date{\today}

\begin{abstract}
 {\it Bound states in the continuum} (BICs) are non-radiating solutions of the wave equation with a spectrum embedded in the continuum of propagating waves of the surrounding space. The complete decoupling of BICs from the radiation continuum makes their excitation impossible from the far-field. Here, we develop a general theory of parametric excitation of BICs in nonlinear systems with Kerr-type nonlinearity via spontaneous symmetry breaking, which results in a coupling of a BIC and a bright mode of the system. Using the temporal coupled-mode theory and perturbation analysis, we found the threshold intensity for excitation of a BIC and study the possible stable and unstable solutions depending on the pump intensity and frequency detuning between the pump and BIC. We revealed that at some parameters of the pump beam,  there are no stable solutions and the BIC can be used for frequency comb generation. Our findings can be very promising for use in nonlinear photonic devices and all-optical networks.
\end{abstract}

\maketitle

\section{Introduction}
{\it Bound state in the continuum} (BIC) is a non-radiating solution with an energy embedded in the continuum spectrum of propagating modes of the surrounding space. They were first revealed in quantum mechanics in 1929 by J. von Neumann and E. Wigner~\cite{vonneumann1993ueber}. Despite the proposed idea has been never implemented in practice for quantum-mechanical systems, it affects the development of atomic physics~\cite{stillinger1975bound, dittes1990avoided}, acoustics~\cite{parker1967resonance,koch1983resonant,evans1994existence}, and hydrodynamics~\cite{ursell1951trapping,retzler2001trapped,cobelli2011experimental}. In recent years, BICs attract more and more attention in nanophotonics~\cite{hsu2016bound,koshelev2019nonradiating,marinica2008bound,bulgakov2008bound} providing the drastic enhancement of the electromagnetic field and its localization at the nanoscale. The advantages of BICs were demonstrated for lasing~\cite{hirose2014wattclass,kodigala2017lasing,wu2020room,yang2020low}, sensing~\cite{romano2018label,tittl2018imagingbased}, filtering~\cite{foley2014symmetryprotected,doskolovich2019integrated}, enhancement of light-matter interaction~\cite{koshelev2018strong,kravtsov2020nonlinear}, nonlinear photonics~\cite{bulgakov2014robust,yuan2016diffraction,yuan2017strong,carletti2018giant,bulgakov2019nonlinear,koshelev2019meta,koshelev2020subwavelength,minkov2019doubly,maksimov2020optical,carletti2019high}, and vortex generation~\cite{doeleman2018experimental,wang2020generating}.    

\begin{figure}[htbp]
  \centering
  \includegraphics[width=0.9\linewidth]{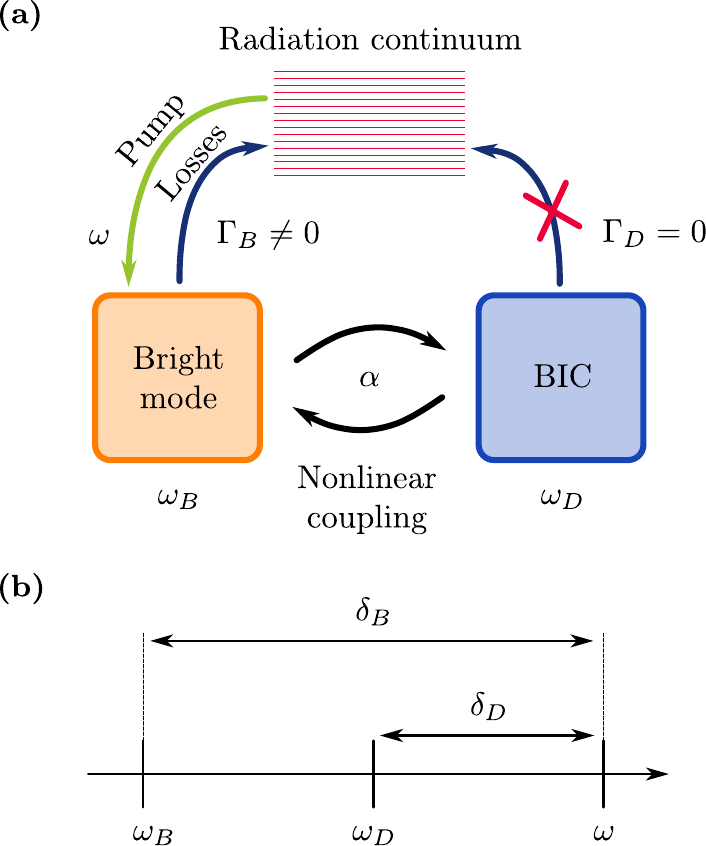}
  \caption{(a) Schematic picture of the considered system consisting of two modes with a nonlinear coupling. One of them is leaky or bright mode, directly coupled to the radiation continuum, and another one is a genuine BIC with no radiative loss. (b) Schematic picture of the relations between eigenfrequencies of the modes $\omega_{B,D}$ and the pump frequency $\omega$.}
  \label{fig:system}
\end{figure}

BICs are formed due to the complete decoupling of leaky modes (resonant states) from the radiating waves. The decoupling can happen, for example, in the system with separable potentials or due to parameter tuning (Fridrich-Wintgen scenario)~\cite{friedrich1985interfering}. More specific mechanisms are discussed in Ref.~\cite{hsu2016bound}. Periodic structures with a rotational symmetry can naturally host so-called {\it symmetry-protected} BICs when the coupling between the radiation continuum and BICs is forbidden in virtue of the symmetry reasons~\cite{hsu2013bloch,zhen2014topological,sadrieva2019multipolar,Overvig2020Selection}. Such states are robust to the presence of a substrate and fabrication inaccuracies preserving the symmetry of the sample.  

The high-Q states are very beneficial, for example, for sensing as they manifest themselves in transmission/reflection spectra as narrow Fano-type resonances whose spectral shift due the presence of an analyte can be precisely measured~\cite{wu2012fano,mouadili2013theoretical,yanik2011seeing}. However, the efficient excitation of high-Q states and strong field enhancement inside the resonator require the fulfillment of the critical coupling condition when the intrinsic losses $\Gamma_{\text{\text{int}}}$ of the resonator are equal to the radiative ones $\Gamma_{\text{\text{rad}}}$~\cite{maier2006plasmonic,seok2011radiation}. Therefore, the efficient excitation of a genuine BIC with $\Gamma_{\text{\text{rad}}}=0$ from the far-field is a challenging problem that can substantially extend its applicability in nanophotonics and metadevices.       

BICs can be excited by near-fields of quantum light sources like quantum wells, quantum dots, NV-centers, dye molecules, etc~\cite{dyakov2020photonic,kodigala2017lasing,zhu2020manipulating,seo2020fourier,yang2020low}. Another way is to provide a finite but tiny coupling between the radiation continuum and BIC. In this case, the BIC turns into a {\it quasi-BIC}. The precise control of radiation losses of quasi-BIC in periodic structures or single resonators can be achieved via the symmetry breaking of the unit cell or the shape of the resonator~\cite{koshelev2018asymmetric,deriy2021bound}. This method is a powerful tool for on-demand engineering of high-Q states.
The structures supporting quasi-BICs have proven themselves well for nonlinear photonics and sensing~\cite{bernhardt2020quasi,zograf2020high,anthur2020continuous,vyas2020improved,chen2020integrated,tseng2020dielectric}. 
Nonlinear polarization induced by the external field can be also used for the excitation of a genuine BIC. Thus, it was shown theoretically that BIC can be excited due to Kerr nonlinearity or second harmonic generation~\cite{yuan2014bilateral,bulgakov2015alloptical,krasikov2018nonlinear,volkovskaya2020multipolar,yuan2020excitation}.    

In this work, we develop a general theory of parametric excitation of BICs in systems with Kerr-type nonlinearity via spontaneous symmetry breaking resulting in a coupling between the BIC and the bright mode.  Figure~\ref{fig:system}(a) illustrates the main idea of the paper.  The BIC at frequency $\omega_D$ has vanishing radiative loss ($\Gamma_D=0$). The bright mode at frequency  $\omega_B$ has a finite radiative loss ($\Gamma_B\neq0$).  The bright mode and BIC usually appear in pairs as symmetric and anti-symmetric solutions of the wave equation (see, e.g., Ref.~\cite{lee2020polarization}). The pump at frequency $\omega$ excites only the bright mode in the linear regime. However, in the nonlinear regime, when the pump intensity is high enough, the division of the solutions into the symmetric (bright)  and  anti-symmetric (BIC) is not applicable as these solutions are no longer robust -- the {\it spontaneous symmetry breaking} occurs~\cite{malomed2013spontaneous,mayteevarunyoo2008spontaneous,kevrekidis2005spontaneous}. It can be interpreted as the nonlinear coupling between the BIC and bright mode resulting in a parametric excitation of the BIC. Therefore, at some pump intensity, the stationary solution corresponding to BIC in the linear case becomes hybrid (without certain parity) and it can be represented as a combination of the bright mode and BIC.
 
The rest of the paper is organized as follows. In Sec.~\ref{sec:model} we describe the physical models and introduce the master equations. Then, in Sec.~\ref{sec:bright} we discuss excitation of BIC in terms of stability of the stationary solutions. After that in Sec.~\ref{sec:hybrid} we describe stationary hybrid states, which occur due to nonlinearity, and finally, in Sec.~\ref{sec:dynamics} we consider dynamics of a nonlinear system and possible operating regimes. Section~\ref{sec:conclusion} summarizes the obtained results and concludes the paper. The main text is supported by Appendix.
In Appendix~\ref{sec:stability}, we provide a detailed description of the stability analysis. Appendix~\ref{sec:phase_portraits} contains the example of trajectories in phase space and graphical representation of the frequency comb generation. Additionally, Supplemental Materials contains the derivation of coupled-mode equations for RLC circuit, which may be used as an experimental model for the verification of the obtained results.

\section{Model}
\label{sec:model}

The following analysis is done in the framework of the coupled-mode theory, which is applicable for a wide range of the system including, for example, atomic systems, resonance photonic and acoustic structures~\cite{yariv1973coupledmode,haus1991coupledmode,ostrovskaya2000coupledmode,fan2003temporal,maksimov2015coupled}. 
One of the simplest mechanisms of BIC formation is based on the coupling between two identical resonances with frequency $\omega_0$ and a finite radiation rate $\gamma_\text{r}$. 
The coupling mixes the resonances, leading to the appearance of two normal modes of different parity (symmetric and asymmetric). When the coupled leaky resonances interfere constructively, they form a symmetric superradiant state with radiation rate $2\gamma_{r}$ -- an analogue of Dicke superradiant state for two emitters~\cite{dicke1954coherence, mlynek2014observation}. The destructive interference cancels out the far-field and, thus, the radiative losses vanish and a symmetry-protected BIC appears.
It is worth mentioning that a symmetry-protected BIC appears exclusively due to the symmetry of the system and the interaction potential allowing the division of the normal modes into symmetric and anti-symmetric.

We consider the  system with a conservative cubic nonlinearity and denote the frequencies of the BIC and bright mode as $\omega_D$ and $\omega_B$, respectively, and their complex amplitudes as $D$ and $B$. The amplitude and the frequency of the pump we denote as $p$ and $\omega$, respectively (see Supplementary Materials for details). The evolution of $B$ and $D$ can be described by the coupled-mode equations written for the slowly varying complex amplitudes of the modes, as in Refs.~\cite{yulin2005dissipative, maksimov2013symmetry} or Supplementary Materials~\footnote{See Supplemental Material for derivation of coupled-modes equations for a system of two nonlinearly coupled RLC circuits.}, for instance:
\begin{widetext}
\centering
\begin{equation}
  \begin{aligned}
    \dot{D} &= - i \delta_D D - \Gamma_D D + i \alpha D \left(|D|^2 + 2 |B|^2\right) + i \alpha B^2 D^*,
    \\
    \dot{B} &= - i \delta_B B - \Gamma_B B + i \alpha B \left(|B|^2 + 2 |D|^2\right) + i \alpha D^2 B^* + p.
  \end{aligned}
  \label{eq:CME_BD}
\end{equation}
\end{widetext}
All the parameters and variables in this system are dimensionless and their normalization is discussed in Supplementary Materials. 
Here $\delta_D$, $\delta_B$ are the detunings of the BIC and the bright mode from the pump frequency $\omega$ [see Fig.~\ref{fig:system}(b)]. Note, that in these equations we assume that the time dependence is $e^{i \omega t}$.
The coefficient of nonlinear coupling $\alpha$ is responsible for the possible symmetry breaking and coupling between the BIC and bright mode.  In the general case both equations for $D$ and $B$ have the decay coefficients $\Gamma_B$ and $\Gamma_D$, respectively. The coefficient $\Gamma_D$ accounts for the non-radiative loss of BIC, which is almost unavoidable in practice, and $\Gamma_B$ contains both the radiative and non-radiative loss of the bright mode (see Supplementary Materials for details), and for quasi-BIC $\Gamma_{B} \gg \Gamma_{D}$.

As we mentioned above, such a system of equations is quite generic and, thus, the further results  are applicable for a broad range of systems. However, to have an illustrative practical example, we suggest in Supplementary Materials a particular system of coupled RLC circuits exactly described by Eqs.~\eqref{eq:CME_BD}. The choice of the system is based on the fact that electrical circuits proved themselves to be a good platform for observation of BICs~\cite{li2020bound}.
The absence of pump term in the equation for the BIC clearly indicates that it is decoupled from the far-field. 
However, the nonlinear term $i \alpha B^2 D^*$ can be considered as an effective parametric amplification (driving force) for the BIC. The excitation is possible only at specific values of the pump amplitude $p$ and frequency detuning $\delta_D$, for which the solution $D=0$ becomes unstable and, thus, any fluctuation in $D$ results in the development of instability -- parametric excitation of BIC. 

\section{Conditions of BIC excitation}~\label{sec:bright}

\begin{figure}[t]
  \centering
  \includegraphics[width=\linewidth]{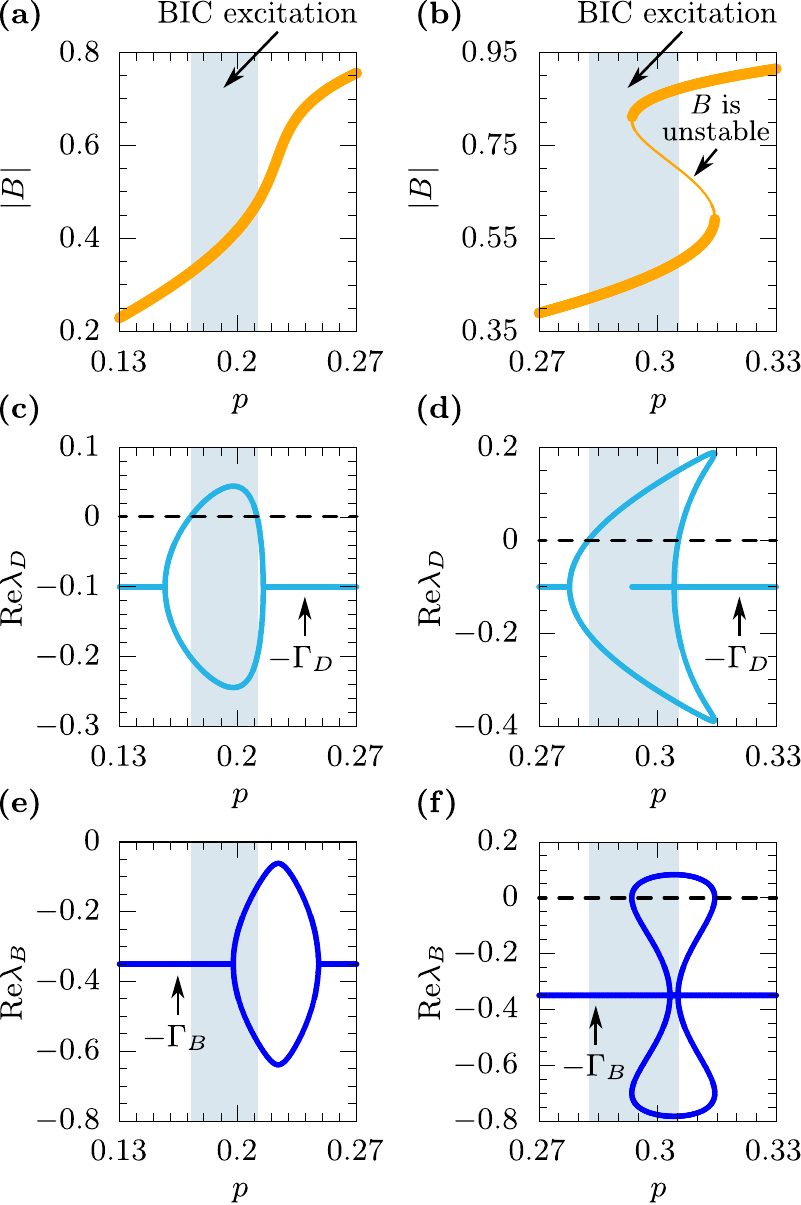}
  \caption{(a, b) Stationary solutions of Eq.~\eqref{eq:CME_B0}, when $D~=~0$. Thick dots correspond to stable solutions for which $\mathrm{Re}~\lambda_B~<~0$ and thin dots correspond to unstable solutions for which  $\mathrm{Re} \lambda_B \geq 0$. Plot of $\mathrm{Re} \lambda_D$ vs $p$ is shown in panels (c, d). Plot of $\mathrm{Re} \lambda_D$ vs $p$ is shown in panels (e, f). Shadowed areas indicate the pump ranges at which the BIC is excited, i.e. where $\mathrm{Re} \lambda_D>0$. The dependencies are plotted for the following parameters:  $\Gamma_D = 0.1$, $\Gamma_B = 0.35$, $\delta_B = \delta_D + 0.25$ and $\alpha = 1$. Detuning $\delta_D$ is equal to $0.25$ for (a,c,e) and $0.5$ for (b,d,f). 
 }
  \label{fig:B0_stability}
\end{figure}

To find the parameters at which the excitation of BIC is possible we need to analyze the stability of the solution $D(\tau)=0$ of Eqs.~\eqref{eq:CME_BD}. This system for the coupled modes reduces to the single equation if we put $D = 0$:
\begin{equation}
	\dot{B} = - i \delta_B B - \Gamma_B B + i \alpha B |B|^2  + p.
	\label{eq:CME_B0}
\end{equation}
The stationary solutions of this equations can be easily found setting $\dot{B}=0$. The amplitude $|B|$ as a function of the pump $p$ for various detuning parameters $\delta_D$ is shown in Figs.~\ref{fig:B0_stability}(a) and \ref{fig:B0_stability}(b). All the parameters are listed in the caption. One can see that the increase of the detuning $\delta_B$ results in the appearance of S-shaped dependence characteristic for non-linear systems.  

To analyze the stability of the found stationary solutions corresponding to the BIC and bright mode we will follow the standard procedure adding fluctuations to them (see Appendix~\ref{sec:stability} for details):
\begin{equation}
	\begin{aligned}
	    \delta B(\tau) &= (m_1 e^{\lambda_B \tau} + n_1 e^{\lambda_B^* \tau}) e^{i \textcolor{black}{\delta_D} \tau},
	    \\
		\delta D (\tau) &= (m_2 e^{\lambda_D \tau} + n_2 e^{\lambda_D^* \tau}) e^{i \delta_D \tau}.
  \end{aligned}
\end{equation}
Here $m_{1,2}$ and $n_{1,2}$ are small arbitrary functions. The system of Eqs.~\eqref{eq:CME_BD} can be linearized with respect to these small functions and reduced to the homogeneous form. The complex parameters $\lambda_D$ and $\lambda_B$ are found from the condition of solvability of the linearized homogeneous system. The exact expressions for $\lambda_{B}$ and $\lambda_{D}$ can be written as (see Appendix~\ref{sec:stability}):
  \begin{align}
   & \lambda_i = - \Gamma_i \pm \sqrt{- \delta_i^2 + 4 \alpha \delta_i |B|^2 - 3 \alpha^2 |B|^4}. 
  \label{eq:lambdas}
  \end{align}
Here $i=B,D$. The stabiltiy of the solutions $B(\tau)$ and $D(\tau)$ is defined by the sign of the real part of $\lambda_{B}$ and $\lambda_{D}$, respectively. If it is positive then the corresponding solution is unstable and vice versa. Therefore, there are four possible scenarios depending on the detuning $\delta_B$, loss $\Gamma_B$, and pump $p$. All of the them are illustrated in Fig.~\ref{fig:B0_stability}. The left columns  [panels (a), (c), and (e)] correspond to the case when $\text{Re}\lambda_B<0$ for all considered values of the pump. Nevertheless, there is a range of $p$ (shaded region), where BIC is excited. 

The right column of Fig.~\ref{fig:B0_stability} [panels (b), (d), and (f)] corresponds to the case when the bright mode demonstrates a bistable behaviour. The middle part of the S-shaped curve in Fig.~\ref{fig:B0_stability}(b) plotted with a thin solid line is unstable. One can see that BIC can be excited for both cases, when $\text{Re}\lambda_B>0$ or $\text{Re}\lambda_B<0$. It worth mentioning that $\Gamma_i$ directly affects the stability of the corresponding mode and there are no unstable solutions in a system with high losses.

\begin{figure}[htbp]
  \centering
  \includegraphics[width=\linewidth]{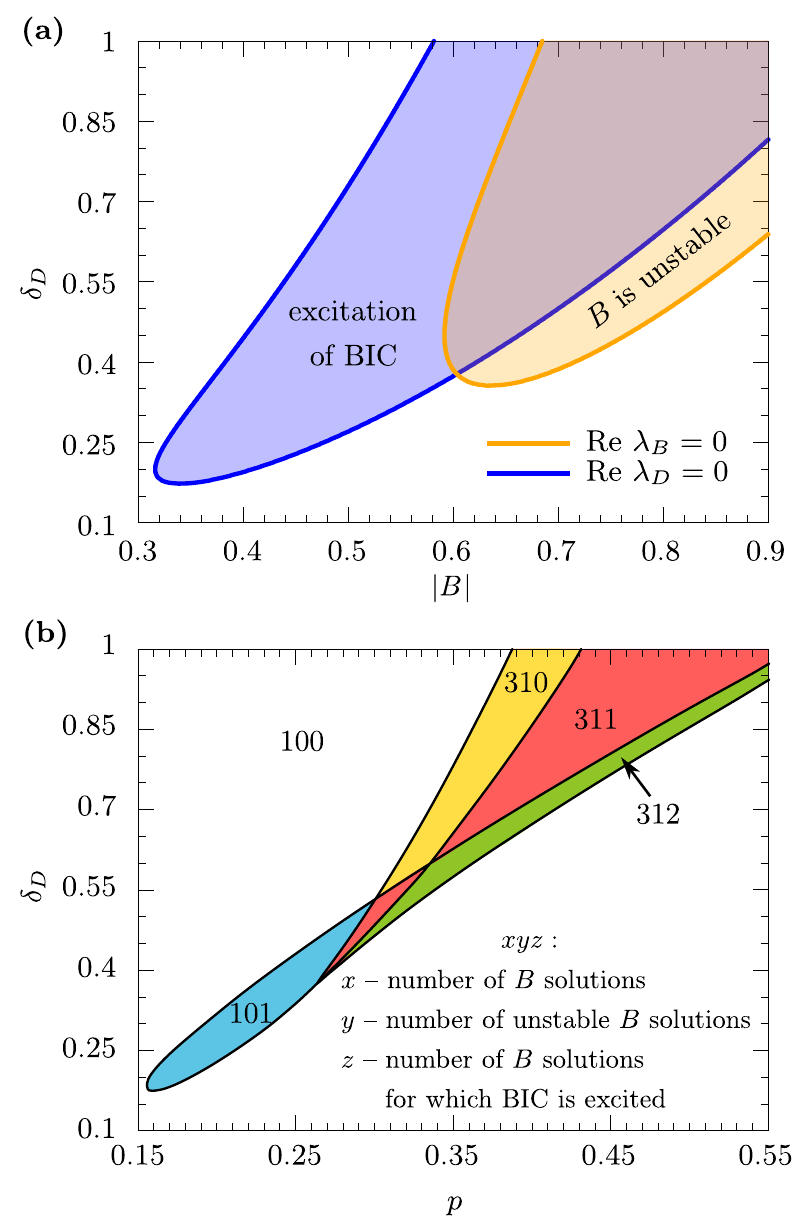}%
  \caption{ (a) Arnold tongues (shaded areas) indicate the regions, where the bright mode is unstable and the parametric excitation of BIC is possible. These regions are defined by the inequalities~\eqref{eq:D_excitation}. Inside the blue shaded area the solution $D = 0$ is unstable, i.e. $\mathrm{Re} \lambda_D > 0$, and the BIC can be excited. Similarly, the shaded orange area indicates the instability of the bright mode. The diagram (b) shows possible regimes of the bright mode stability and BIC excitation. Numerical $xyz$-code has the following meaning: $x$ stands for the total number of the bright mode solutions, $y$ shows how much of them are unstable (i.e. $\mathrm{Re} \lambda_B > 0$) and $z$ indicates the number of bright solutions for which excitation of BIC is possible (i.e. $\mathrm{Re} \lambda_D > 0$). For example, $312$ means that there are $3$ bright solutions, $1$ of them is unstable, and $2$ of them leads to excitation of BIC. Each graph in this figure was plotted for $\Gamma_D = 0.1$, $\Gamma_B = 0.35$, $\delta_B = \delta_D + 0.25$ and $\alpha = 1$.}
  \label{fig:B0_diagram}
\end{figure}

The equation~\eqref{eq:lambdas} allows to find the range of amplitudes $|B|$ and frequency detunings $\delta_{B,D}$ for which the BIC is excited and the bright mode is unstable. These ranges are defined by the following inequalities:
\begin{equation}
    \frac{2 \delta_i - \sqrt{\delta_i^2 - 3 \Gamma_i^2}}{3\alpha}  < |B|^2 < \frac{2 \delta_i + \sqrt{\delta_i^2 - 3 \Gamma_i ^2}}{3\alpha}.
  \label{eq:D_excitation}
\end{equation}
Here $i=B,D$. The graphical representations of these inequalities is shown in Fig.~\ref{fig:B0_diagram}(a). In the shaded areas, also known as {\it Arnold tongues}, the solutions are unstable. These graphs explicitly demonstrate that excitation of BIC at high amplitudes of the bright mode requires high detunings $\delta_D$, since the frequency of the BIC shifts with the increase of $|B|$. In case of the negative Kerr shift, such a detuning would be negative. The inequality~\eqref{eq:D_excitation} allows to find the maximal loss $\Gamma_{B,D}$ as a function $\delta_{B,D}$ at which the parametric excitation of BIC and instability the bright mode is possible:
\begin{equation}
    \Gamma_{B,D} \leq \delta_{B,D}/\sqrt{3}.
\end{equation}
Importantly, if $\Gamma_D = 0$, the tip of the Arnold tongue touches the point $|B| = 0$ and $\delta_D = 0$ [see  Eq.~\eqref{eq:D_excitation}]. Therefore, the parametric excitation of BIC is thresholdless, i.e. it is possible at any arbitrary small values of the pump amplitude $p$. On the other hand, $\delta_B > \delta_D$ according to the chosen definition [see Fig.~\ref{fig:system}(b)], so it is always non-zero, since $\delta_D \geq 0$ according to the structure of Eqs.~\eqref{eq:CME_BD}. Therefore, the bright mode becomes unstable only if the pump amplitude exceeds some threshold value. 

The diagram in Fig.~\ref{fig:B0_diagram}(b) summarizes possible stability regimes of the bright mode and the BIC solutions assuming variation of the pump amplitude $p$ and the detuning $\delta_D$. These regimes are labeled by a numerical $xyz$-code, where $x$ is the total number of solutions, $y$ shows how many of these solutions are unstable and $z$ stays for the number of solutions for which excitation of BIC is possible, i.e. when $\mathrm{Re} \lambda_D > 0$. For example, the region labeled as $100$ in Fig.~\ref{fig:B0_diagram}(b) indicates that Eqs.~\eqref{eq:CME_BD} has only one solution for $B$ which is stable and the BIC is not excited. Similarly, the region labeled as $312$ in the same diagram corresponds to the case when Eqs.~\eqref{eq:CME_BD} has three solutions for $B$, one of them is unstable (bistability), and the parametric excitation of the BIC takes place for two of the solutions.

Thus, we analysed the conditions necessary for the excitation of the BIC. The next question that we are going to answer: are there any stable states after BIC excitation? 

\section{Hybrid States}~\label{sec:hybrid}
In Sec.~\ref{sec:bright} we analyzed the solutions of Eqs.~\eqref{eq:CME_BD} for which $D(\tau) = 0$. However, these are not all stationary solutions. When the symmetry breaks, the division of the solutions into the symmetric and anti-symmetric is no longer applicable and stationary hybrid states can appear. For the hybrid states, both $B$ and $D$ are not zero. To find all possible stationary hybrid solution we need to set all time derivatives in Eqs.~\eqref{eq:CME_BD} without assumption that $D(\tau) = 0$.       

\begin{figure}[htbp!]
  \centering
  \includegraphics[width=\linewidth]{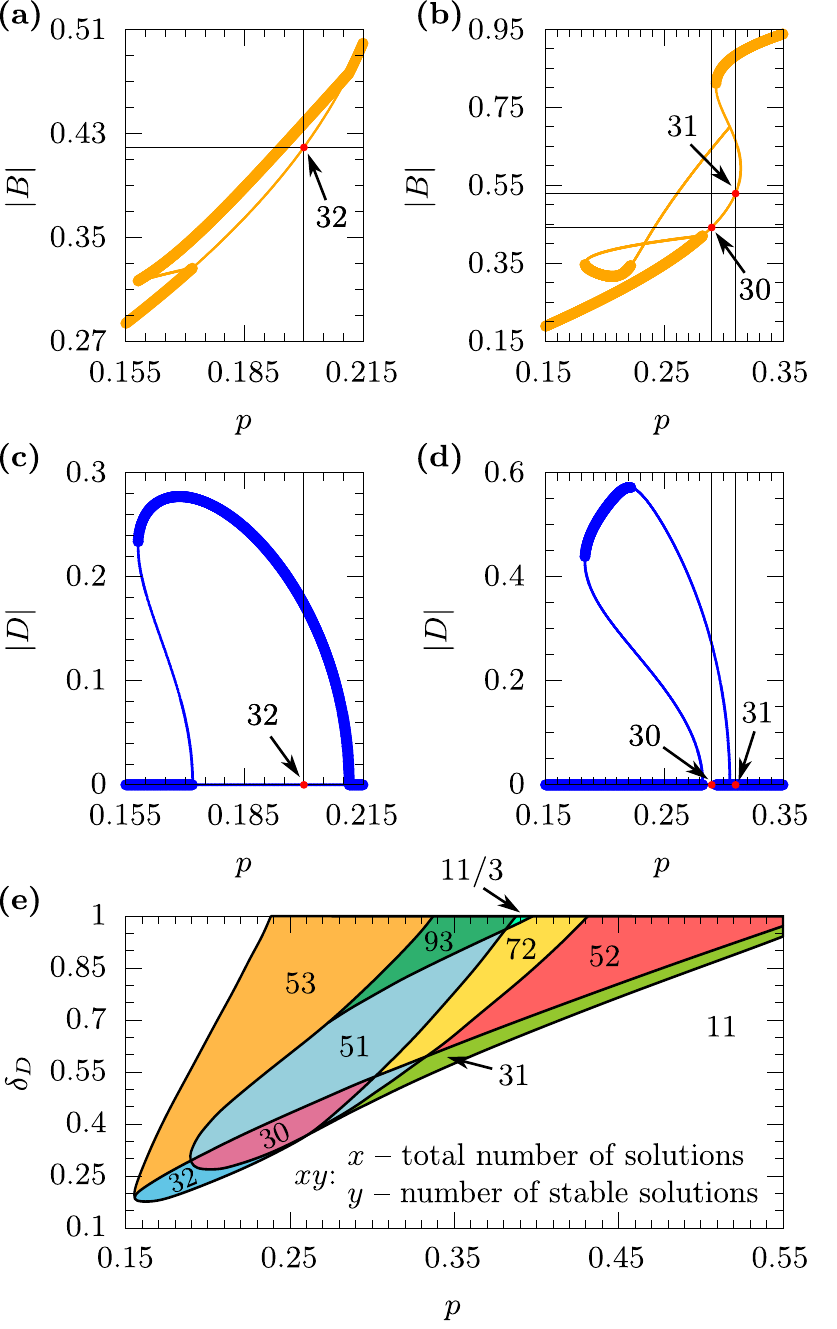}%
  \caption{Stationary solutions represented as the dependence of $|B|$ and $|D|$ amplitudes on the amplitude of pump $p$. Thick dots represents the stable solutions and thin dots are for unstable solutions. Detuning is $\delta_D = 0.25$ for (a,c) and $\delta_D = 0.5$ for (b,d). The diagram (e) shows possible stability regimes of the stationary solutions. Numerical $xy$-code has the following meaning: $x$ stays for the total number of solutions and $y$ shows how much of these solutions are stable. Solutions $D \neq 0$ are doubly degenerate, which is accounted by the number of solutions. The dependencies are plotted for the following parameters: $\Gamma_D = 0.1$, $\Gamma_B = 0.35$, $\alpha = 1$ and $\delta_B = \delta_D + 0.25$.}
  \label{fig:BD_bistability}
\end{figure}

Figures~\ref{fig:BD_bistability}(a)--\ref{fig:BD_bistability}(d) show the dependence of the amplitudes $|B|$ and $|D|$ on the pump $p$ for the stationary solutions of Eqs.~\eqref{eq:CME_BD} accounting the hybrid states. The dependencies in panels (a,c) and (b,d) are plotted for the same parameters as in Figs.~\ref{fig:B0_stability}(a) and \ref{fig:B0_stability}(b), respectively. One can see that for some solutions both amplitudes $B$ and $D$ are not zero and, thus, the states are hybrid.  The stable solutions are plotted with thick lines and the unstable solutions are plotted with thin lines.    

\begin{figure}[t]
  \centering
  \includegraphics[width=\linewidth]{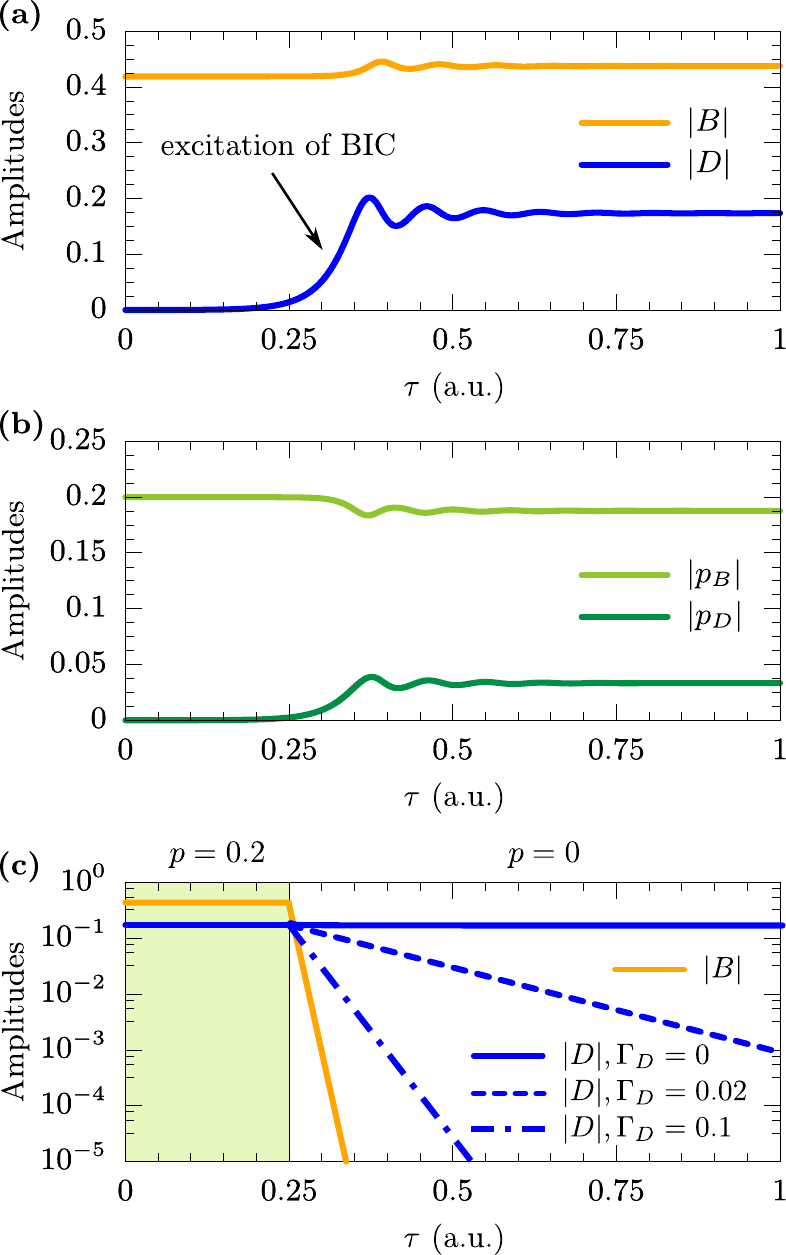}
  \caption{(a) Stable hybrid solution, which exist in the system constantly pumped with amplitude $p = 0.2$. Here $\Gamma_D = 0.1$. (b) Corresponding values of $|p_{B,D}|$ shows that the effective driving force acting on the modes actually may change with time due to the nonlinear coupling even if the pump amplitude is constant. (c) Demonstration of the energy locking. The system initially is in stable hybrid state, constantly pumped with amplitude $p = 0.2$ (shaded area). Then at time $\tau = 0.25$ the pump is switched off, so the bright mode amplitude quickly decreases to zero. The BIC can remain in the system for a longer period of time, defined by the decay rate $\Gamma_D$, which accounts for non-radiative losses. All graphs in this figure were obtained for $\Gamma_B = 0.35$, $\alpha = 1$, $\delta_D = 0.25$ and $\delta_B = 0.5$.}
  \label{fig:D_stable}
\end{figure}

One can see that Eqs.~\eqref{eq:CME_BD} have a simple form but their solutions are very manifold. The diagram in Fig.~\ref{fig:BD_bistability}(e) summarizes possible stability regimes of the considered system depending on the pump amplitude $p$ and the detuning $\delta_D$. Each region is labelled by a numerical $xy$-code, where $x$ is the total number of solutions and $y$ shows how many of them are stable. For example, in the region labelled as $32$ there are three solutions and two of them are stable. Note, that hybrid states are doubly degenerate, thus, they are counted twice in the number of solutions. This degeneracy arises from the symmetry of a structure as the coupled oscillators are identical in the considered case. When the oscillators are not identical (for example, in photonics this corresponds to the case of an asymmetric unit cell~\cite{koshelev2018asymmetric}), these states may have different energies and the degeneracy lifts off. In particular, for the RLC circuits considered  in Supplementary Materials this means that elements of the coupled circuits have different parameters.

The stability analysis of the hybrid solutions does not reveal the process of their formation. To gain deeper inside into this problem we consider the time-dynamics of the BIC excitation and multi-stability of the bright mode.




\begin{figure*}[t]
    \centering
    \includegraphics[width=\linewidth]{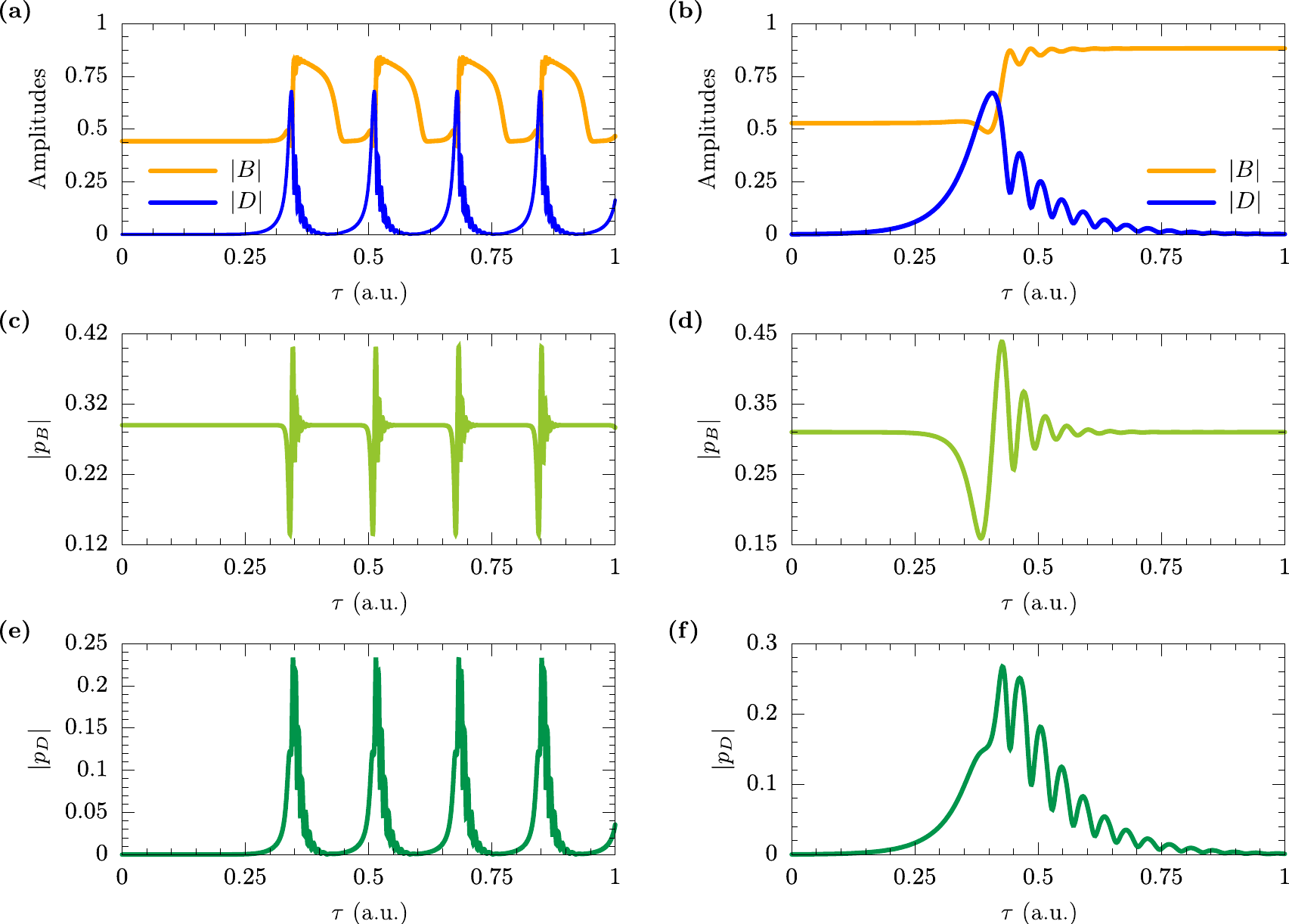}
    \caption{(a) Time-dynamics of the self-oscillatory regime occurring in a system pumped with constant amplitude $p = 0.29$, and (c), (e) corresponding graphs for the driving forces. (b) Single-pulse excitation of BIC. Amplitude of the pump remains constant, $p = 0.31$. Panels (d) and (f) shows corresponding dynamics of the driving forces $p_{B,D}$ . Each graph in this figure was obtained for $\Gamma_D = 0.1$, $\Gamma_B = 0.35$, $\alpha = 1$, $\delta_D = 0.5$ and $\delta_B = \delta_D + 0.25$.}
    \label{fig:BD_self}
\end{figure*}

\section{Time Dynamics \label{sec:dynamics}}
\subsection{Excitation of BIC}


As we mentioned above, the parametric excitation of the BIC is possible due to the spontaneous symmetry breaking when the amplitude $B$ of the bright mode reaches a threshold value at which the division of the solutions into symmetric and anti-symmetric is not energetically favorable. To demonstrate this effect numerically we should add a weak perturbation to the initial condition $D(\tau)=0$. Equations~\eqref{eq:CME_BD} are solved numerically by a standard method implemented in Python using SciPy library. For the crosscheck, the same solutions were obtained with the help of standard ODE solver implemented in MATLAB\textsuperscript{\small\textregistered}.

 

As an example, we took the following parameters: $p=0.2$, $\delta_D=0.25$, $\delta_B=0.5$, $\Gamma_D = 0.1$, $\Gamma_B = 0.35$, $\alpha=1$. According to the analysis from Sec.~\ref{sec:hybrid},  these parameters correspond to the region labelled as 32 in Fig.~\ref{fig:BD_bistability}(e), i.e. when three solutions exist: one unstable pure bright solution and one stable doubly degenerate hybrid solution [see the red points in Figs.~\ref{fig:BD_bistability}(a) and \ref{fig:BD_bistability}(c) labelled as 32].

The time-dynamics of $B$ and $D$ is shown in Fig.~\ref{fig:D_stable}(a). Initially, the system is in a state when $|D|=0$ and $|B|$ is a constant but this state is unstable and the introduced fluctuation in $D$ becomes to develop -- BIC is excited. We can see that the system tends to another stationary stable solution, which is hybrid. Thus, we can conclude that the hybrid solution is an attractor. Indeed, the phase portrait demonstrated in Appendix~\ref{sec:phase_portraits} [see Fig.~\ref{fig:phase_portraits}(a)] reveals the presence of a focus point, associated with the stable hybrid solution.

The energy exchange between the BIC and bright mode can be interpreted in terms of the nonlinear driving forces, which can be written as
  \begin{align}
    p_D &= i \alpha B^2 D^*, \label{eg:driving_force_D}
    \\
    p_B &= i \alpha D^2 B^* + p.
  \end{align}
Figure~\ref{fig:D_stable}(b) demonstrates that the change in amplitudes $B$ and $D$ is induced by these corresponding driving forces $p_B$ and $p_D$. The amplitude $D$ of the BIC starts to grow with the increase of $p_D$ caused by the coupling with the bright mode.

\subsection{Trapping of energy by BIC}

In open linear systems, the coupling matrix is symmetric~\textcolor{black}{\cite{suh2004temporal}}. This means that the coupling and decoupling rates are equal to each other. In particular, an infinite radiative lifetime of a photon in a resonator implies a complete decoupling of the resonator's mode from the radiative continuum that makes it impossible to excite the mode from the far-field. However, in nonlinear systems, this issue can be overcome as the coupling of the mode with the radiative continuum depends on the pump intensity. In our case, the coupling of BIC to the continuum is governed by the amplitude of the bright mode $B$ [see Eqs.~\eqref{eq:CME_BD}]. This allows not only to excite BIC but also trap energy inside the system for a long time~\textcolor{black}{\cite{bulgakov2015alloptical}}. Indeed, let's imagine that the BIC is excited and the system is in a hybrid stationary state supported by an external pump. After the pump is switched off, the amplitude $B$ decreases fast due to the radiation of the bright mode. It 
results in vanishing coupling of BIC to the radiation continuum 
as it is proportional to the coupling term in Eqs.~\eqref{eq:CME_BD} proportional to $B^2$. Therefore, the energy initially pumped to BIC is trapped.

Figure~\ref{fig:D_stable}(c) clearly demonstrates the behaviour of the BIC and the bright mode in this scenario. The system is set into a stationary hybrid state supported by a permanent external pump with amplitude $p=0.2$. One can see that after switch-off of the pump at $\tau=0.25$, the amplitude of the bright mode $B$  decays rapidly but the amplitude of the BIC becomes constant in the absence of non-radiative loss ($\Gamma_D=0$), thus, the energy is locked in the BIC. In the lossy case, the storage time is completely defined by $\Gamma_D$ [see dashed and dot-dashed lines in Fig.~\ref{fig:D_stable}(c)]. This effect may be used for the realization of optical memory~\cite{bulgakov2015alloptical,lannebere2015optical}.

\subsection{Regime of self-oscillations}
It is interesting to note that the instability of the bright state can lead to the dynamical regime of self-sustained oscillations. This happens when the system has no stable states at all. An example of such a region is labelled as $30$ in Fig.~\ref{fig:BD_bistability}(e). Indeed, one can also see from Fig.~\ref{fig:BD_bistability}(b) that the black vertical line $p=0.29$ does not cross the thick orange curve, thus, there are no unstable solutions for such a pump amplitude. Similar behaviour was also discussed in Ref.~\cite{maksimov2013symmetry}.

The time dynamics of $|B|$ and $|D|$ for this case is shown in Fig.~\ref{fig:BD_self}(a). The parameters of the system are listed in the caption. The system is initially in an unstable bright state. Then the instability is developed and the amplitude of the BIC grows reaching some maximal value. However, after that the system returns back to the state close to the initial bright solution, since the effective driving force $p_D$ decreases to zero. Then the process repeats, so the system oscillates as Fig.~\ref{fig:BD_self}(a) shows. 
This behaviour of the system also can be traced with the phase portrait presented in Fig.~\ref{fig:phase_portraits}(b) in Appendix~\ref{sec:phase_portraits}. The initial point corresponding to $D \approx 0$ 
is actually a saddle point. In this case the system is stable with respect to fluctuations in $B$, but unstable with respect to fluctuations in $D$. Therefore, the trajectory near this point is hyperbolic. Moreover, system follows the closed path since the effective driven forces $p_B$ and $p_D$ also demonstrate a periodic behavior [see Fig.~\ref{fig:BD_self}(c) and ~\ref{fig:BD_self}(e)].
Each period of such \textit{self-oscillations} is associated with the presence of a large number of equidistant harmonics [see Fig.~\ref{fig:phase_portraits}(d) in Appendix~\ref{sec:phase_portraits}], which is a manifestation of a frequency comb generation. The idea to use BIC in systems with Kerr nonlinearity for frequency comb generation is also discussed  in Ref.~\cite{pichugin2015frequency}.


Let us also remark that the transition from the unstable pure bright state to the stable pure bright state can be accompanied by the excitation of BIC. The time dynamics of  $|B|$ and $|D|$ for this case is shown in Fig.~\ref{fig:BD_self}(b). The instability  results in growth of $D$ but the rise of the bright mode amplitude $B$ decreases  the effective gain for the BIC [see Fig.~\ref{fig:BD_self}(d) and~\ref{fig:BD_self}(f)]. As the result, the system switches from the unstable to the stable bright state and during this transition the BIC is excited. This is the case labeled as $31$ in Fig.~\ref{fig:BD_bistability}. The phase portrait in this case represents the movement of the system from an unstable node to a stable node, as Fig.~\ref{fig:phase_portraits}(c) in Appendix~\ref{sec:phase_portraits} demonstrates.

\section{Conclusion}~\label{sec:conclusion}
To conclude, we have developed a theory describing the parametric excitation of symmetry-protected BICs in nonlinear systems with Kerr-type nonlinearity due to spontaneous symmetry breaking resulting in a coupling of BIC to the bright mode of the system. The proposed theory is very general. It is based on the two coupled oscillator model that can be applied to a variety of physical systems in optics, acoustic, hydrodynamics, quantum mechanics, etc. The obtained results are especially important for photonics as they provide useful guidelines for the excitation of BICs from the far-field. We also provided a detailed analysis of possible stable and unstable solutions accounting for hybrid states that appear as a result of coupling between the BIC and bright mode. We revealed several interesting regimes of the nonlinear system supporting BIC. In particular, we have shown that the energy from the far-field can be trapped by BIC for a long time, which is limited only by nonradiative losses of BIC. We also have found regimes, when there are no stable solutions at all. The numerical analysis of the time dynamics shows that such regimes can be potentially used for the frequency combs or super-continuum generation. We believe that our findings can help to extend the practical applicability of BIC in different systems.  Especially, the obtained results  can be very promising for nonlinear photonics and all-optical networks.

\begin{acknowledgements}
This work is supported by the Russian Science Foundation (project 18-72-10140). A.B. acknowledges the BASIS foundation and the Grant of the President of the Russian Federation (MK-2224.2020.2). The authors thank Constantin Simovski, Ivan Iorsh, and Dmitrii Maksimov for fruitful discussions.

\end{acknowledgements}

A.C. and S.K. contributed equally to this work. 

\bibliography{bibliography}

\begin{thebibliography}{80}%
\makeatletter
\providecommand \@ifxundefined [1]{%
 \@ifx{#1\undefined}
}%
\providecommand \@ifnum [1]{%
 \ifnum #1\expandafter \@firstoftwo
 \else \expandafter \@secondoftwo
 \fi
}%
\providecommand \@ifx [1]{%
 \ifx #1\expandafter \@firstoftwo
 \else \expandafter \@secondoftwo
 \fi
}%
\providecommand \natexlab [1]{#1}%
\providecommand \enquote  [1]{``#1''}%
\providecommand \bibnamefont  [1]{#1}%
\providecommand \bibfnamefont [1]{#1}%
\providecommand \citenamefont [1]{#1}%
\providecommand \href@noop [0]{\@secondoftwo}%
\providecommand \href [0]{\begingroup \@sanitize@url \@href}%
\providecommand \@href[1]{\@@startlink{#1}\@@href}%
\providecommand \@@href[1]{\endgroup#1\@@endlink}%
\providecommand \@sanitize@url [0]{\catcode `\\12\catcode `\$12\catcode
  `\&12\catcode `\#12\catcode `\^12\catcode `\_12\catcode `\%12\relax}%
\providecommand \@@startlink[1]{}%
\providecommand \@@endlink[0]{}%
\providecommand \url  [0]{\begingroup\@sanitize@url \@url }%
\providecommand \@url [1]{\endgroup\@href {#1}{\urlprefix }}%
\providecommand \urlprefix  [0]{URL }%
\providecommand \Eprint [0]{\href }%
\providecommand \doibase [0]{http://dx.doi.org/}%
\providecommand \selectlanguage [0]{\@gobble}%
\providecommand \bibinfo  [0]{\@secondoftwo}%
\providecommand \bibfield  [0]{\@secondoftwo}%
\providecommand \translation [1]{[#1]}%
\providecommand \BibitemOpen [0]{}%
\providecommand \bibitemStop [0]{}%
\providecommand \bibitemNoStop [0]{.\EOS\space}%
\providecommand \EOS [0]{\spacefactor3000\relax}%
\providecommand \BibitemShut  [1]{\csname bibitem#1\endcsname}%
\let\auto@bib@innerbib\@empty
\bibitem [{\citenamefont {{von Neumann}}\ and\ \citenamefont
  {Wigner}(1993)}]{vonneumann1993ueber}%
  \BibitemOpen
  \bibfield  {author} {\bibinfo {author} {\bibfnamefont {J.}~\bibnamefont {{von
  Neumann}}}\ and\ \bibinfo {author} {\bibfnamefont {E.~P.}\ \bibnamefont
  {Wigner}},\ }in\ \href {\doibase 10.1007/978-3-662-02781-3_19}
  {{\selectlanguage {german}\emph {\bibinfo {booktitle} {{The Collected Works
  of Eugene Paul Wigner: Part A: The Scientific Papers}}}}},\ \bibinfo {series
  and number} {{The Collected Works of Eugene Paul Wigner}},\ \bibinfo {editor}
  {edited by\ \bibinfo {editor} {\bibfnamefont {A.~S.}\ \bibnamefont
  {Wightman}}}\ (\bibinfo  {publisher} {{Springer}},\ \bibinfo {address}
  {{Berlin}},\ \bibinfo {year} {1993})\ pp.\ \bibinfo {pages}
  {291--293}\BibitemShut {NoStop}%
\bibitem [{\citenamefont {Stillinger}\ and\ \citenamefont
  {Herrick}(1975)}]{stillinger1975bound}%
  \BibitemOpen
  \bibfield  {author} {\bibinfo {author} {\bibfnamefont {F.~H.}\ \bibnamefont
  {Stillinger}}\ and\ \bibinfo {author} {\bibfnamefont {D.~R.}\ \bibnamefont
  {Herrick}},\ }\href {\doibase 10/fspg5g} {\bibfield  {journal} {\bibinfo
  {journal} {Phys. Rev. A}\ }\textbf {\bibinfo {volume} {11}},\ \bibinfo
  {pages} {446} (\bibinfo {year} {1975})}\BibitemShut {NoStop}%
\bibitem [{\citenamefont {Dittes}\ \emph {et~al.}(1990)\citenamefont {Dittes},
  \citenamefont {Cassing},\ and\ \citenamefont {Rotter}}]{dittes1990avoided}%
  \BibitemOpen
  \bibfield  {author} {\bibinfo {author} {\bibfnamefont {F.-M.}\ \bibnamefont
  {Dittes}}, \bibinfo {author} {\bibfnamefont {W.}~\bibnamefont {Cassing}}, \
  and\ \bibinfo {author} {\bibfnamefont {I.}~\bibnamefont {Rotter}},\
  }\href@noop {} {\bibfield  {journal} {\bibinfo  {journal} {Z. Phys. A}\
  }\textbf {\bibinfo {volume} {337}},\ \bibinfo {pages} {243} (\bibinfo {year}
  {1990})}\BibitemShut {NoStop}%
\bibitem [{\citenamefont {Parker}(1967)}]{parker1967resonance}%
  \BibitemOpen
  \bibfield  {author} {\bibinfo {author} {\bibfnamefont {R.}~\bibnamefont
  {Parker}},\ }\href {\doibase 10/bz9mb2} {\bibfield  {journal} {\bibinfo
  {journal} {J. Sound Vibration}\ }\textbf {\bibinfo {volume} {5}},\ \bibinfo
  {pages} {330} (\bibinfo {year} {1967})}\BibitemShut {NoStop}%
\bibitem [{\citenamefont {Koch}(1983)}]{koch1983resonant}%
  \BibitemOpen
  \bibfield  {author} {\bibinfo {author} {\bibfnamefont {W.}~\bibnamefont
  {Koch}},\ }\href {\doibase 10/fs2tht} {\bibfield  {journal} {\bibinfo
  {journal} {J. Sound Vibration}\ }\textbf {\bibinfo {volume} {88}},\ \bibinfo
  {pages} {233} (\bibinfo {year} {1983})}\BibitemShut {NoStop}%
\bibitem [{\citenamefont {Evans}\ \emph {et~al.}(1994)\citenamefont {Evans},
  \citenamefont {Levitin},\ and\ \citenamefont
  {Vassiliev}}]{evans1994existence}%
  \BibitemOpen
  \bibfield  {author} {\bibinfo {author} {\bibfnamefont {D.~V.}\ \bibnamefont
  {Evans}}, \bibinfo {author} {\bibfnamefont {M.}~\bibnamefont {Levitin}}, \
  and\ \bibinfo {author} {\bibfnamefont {D.}~\bibnamefont {Vassiliev}},\ }\href
  {\doibase 10/fj9mmn} {\bibfield  {journal} {\bibinfo  {journal} {J. Fluid
  Mech.}\ }\textbf {\bibinfo {volume} {261}},\ \bibinfo {pages} {21} (\bibinfo
  {year} {1994})}\BibitemShut {NoStop}%
\bibitem [{\citenamefont {Ursell}(1951)}]{ursell1951trapping}%
  \BibitemOpen
  \bibfield  {author} {\bibinfo {author} {\bibfnamefont {F.}~\bibnamefont
  {Ursell}},\ }\href {\doibase 10/cqpb4z} {\bibfield  {journal} {\bibinfo
  {journal} {Math. Proc. Cambridge Philos. Soc.}\ }\textbf {\bibinfo {volume}
  {47}},\ \bibinfo {pages} {347} (\bibinfo {year} {1951})}\BibitemShut
  {NoStop}%
\bibitem [{\citenamefont {Retzler}(2001)}]{retzler2001trapped}%
  \BibitemOpen
  \bibfield  {author} {\bibinfo {author} {\bibfnamefont {C.~H.}\ \bibnamefont
  {Retzler}},\ }\href {\doibase 10/b68wft} {\bibfield  {journal} {\bibinfo
  {journal} {Appl. Ocean Res.}\ }\textbf {\bibinfo {volume} {23}},\ \bibinfo
  {pages} {249} (\bibinfo {year} {2001})}\BibitemShut {NoStop}%
\bibitem [{\citenamefont {Cobelli}\ \emph {et~al.}(2011)\citenamefont
  {Cobelli}, \citenamefont {Pagneux}, \citenamefont {Maurel},\ and\
  \citenamefont {Petitjeans}}]{cobelli2011experimental}%
  \BibitemOpen
  \bibfield  {author} {\bibinfo {author} {\bibfnamefont {P.~J.}\ \bibnamefont
  {Cobelli}}, \bibinfo {author} {\bibfnamefont {V.}~\bibnamefont {Pagneux}},
  \bibinfo {author} {\bibfnamefont {A.}~\bibnamefont {Maurel}}, \ and\ \bibinfo
  {author} {\bibfnamefont {P.}~\bibnamefont {Petitjeans}},\ }\href {\doibase
  10/ffsrwt} {\bibfield  {journal} {\bibinfo  {journal} {J. Fluid Mech.}\
  }\textbf {\bibinfo {volume} {666}},\ \bibinfo {pages} {445} (\bibinfo {year}
  {2011})}\BibitemShut {NoStop}%
\bibitem [{\citenamefont {Hsu}\ \emph {et~al.}(2016)\citenamefont {Hsu},
  \citenamefont {Zhen}, \citenamefont {Stone}, \citenamefont {Joannopoulos},\
  and\ \citenamefont {Solja{\v{c}}i{\'c}}}]{hsu2016bound}%
  \BibitemOpen
  \bibfield  {author} {\bibinfo {author} {\bibfnamefont {C.~W.}\ \bibnamefont
  {Hsu}}, \bibinfo {author} {\bibfnamefont {B.}~\bibnamefont {Zhen}}, \bibinfo
  {author} {\bibfnamefont {A.~D.}\ \bibnamefont {Stone}}, \bibinfo {author}
  {\bibfnamefont {J.~D.}\ \bibnamefont {Joannopoulos}}, \ and\ \bibinfo
  {author} {\bibfnamefont {M.}~\bibnamefont {Solja{\v{c}}i{\'c}}},\ }\href@noop
  {} {\bibfield  {journal} {\bibinfo  {journal} {Nat. Rev. Mater.}\ }\textbf
  {\bibinfo {volume} {1}},\ \bibinfo {pages} {1} (\bibinfo {year}
  {2016})}\BibitemShut {NoStop}%
\bibitem [{\citenamefont {Koshelev}\ \emph
  {et~al.}(2019{\natexlab{a}})\citenamefont {Koshelev}, \citenamefont
  {Favraud}, \citenamefont {Bogdanov}, \citenamefont {Kivshar},\ and\
  \citenamefont {Fratalocchi}}]{koshelev2019nonradiating}%
  \BibitemOpen
  \bibfield  {author} {\bibinfo {author} {\bibfnamefont {K.}~\bibnamefont
  {Koshelev}}, \bibinfo {author} {\bibfnamefont {G.}~\bibnamefont {Favraud}},
  \bibinfo {author} {\bibfnamefont {A.}~\bibnamefont {Bogdanov}}, \bibinfo
  {author} {\bibfnamefont {Y.}~\bibnamefont {Kivshar}}, \ and\ \bibinfo
  {author} {\bibfnamefont {A.}~\bibnamefont {Fratalocchi}},\ }\href@noop {}
  {\bibfield  {journal} {\bibinfo  {journal} {Nanophotonics}\ }\textbf
  {\bibinfo {volume} {8}},\ \bibinfo {pages} {725} (\bibinfo {year}
  {2019}{\natexlab{a}})}\BibitemShut {NoStop}%
\bibitem [{\citenamefont {Marinica}\ \emph {et~al.}(2008)\citenamefont
  {Marinica}, \citenamefont {Borisov},\ and\ \citenamefont
  {Shabanov}}]{marinica2008bound}%
  \BibitemOpen
  \bibfield  {author} {\bibinfo {author} {\bibfnamefont {D.}~\bibnamefont
  {Marinica}}, \bibinfo {author} {\bibfnamefont {A.}~\bibnamefont {Borisov}}, \
  and\ \bibinfo {author} {\bibfnamefont {S.}~\bibnamefont {Shabanov}},\
  }\href@noop {} {\bibfield  {journal} {\bibinfo  {journal} {Phys. Rev. Lett.}\
  }\textbf {\bibinfo {volume} {100}},\ \bibinfo {pages} {183902} (\bibinfo
  {year} {2008})}\BibitemShut {NoStop}%
\bibitem [{\citenamefont {Bulgakov}\ and\ \citenamefont
  {Sadreev}(2008)}]{bulgakov2008bound}%
  \BibitemOpen
  \bibfield  {author} {\bibinfo {author} {\bibfnamefont {E.~N.}\ \bibnamefont
  {Bulgakov}}\ and\ \bibinfo {author} {\bibfnamefont {A.~F.}\ \bibnamefont
  {Sadreev}},\ }\href@noop {} {\bibfield  {journal} {\bibinfo  {journal} {Phys.
  Rev. B}\ }\textbf {\bibinfo {volume} {78}},\ \bibinfo {pages} {075105}
  (\bibinfo {year} {2008})}\BibitemShut {NoStop}%
\bibitem [{\citenamefont {Hirose}\ \emph {et~al.}(2014)\citenamefont {Hirose},
  \citenamefont {Liang}, \citenamefont {Kurosaka}, \citenamefont {Watanabe},
  \citenamefont {Sugiyama},\ and\ \citenamefont {Noda}}]{hirose2014wattclass}%
  \BibitemOpen
  \bibfield  {author} {\bibinfo {author} {\bibfnamefont {K.}~\bibnamefont
  {Hirose}}, \bibinfo {author} {\bibfnamefont {Y.}~\bibnamefont {Liang}},
  \bibinfo {author} {\bibfnamefont {Y.}~\bibnamefont {Kurosaka}}, \bibinfo
  {author} {\bibfnamefont {A.}~\bibnamefont {Watanabe}}, \bibinfo {author}
  {\bibfnamefont {T.}~\bibnamefont {Sugiyama}}, \ and\ \bibinfo {author}
  {\bibfnamefont {S.}~\bibnamefont {Noda}},\ }\href {\doibase 10/f5zjgt}
  {\bibfield  {journal} {\bibinfo  {journal} {Nat. Photonics}\ }\textbf
  {\bibinfo {volume} {8}},\ \bibinfo {pages} {406} (\bibinfo {year}
  {2014})}\BibitemShut {NoStop}%
\bibitem [{\citenamefont {Kodigala}\ \emph {et~al.}(2017)\citenamefont
  {Kodigala}, \citenamefont {Lepetit}, \citenamefont {Gu}, \citenamefont
  {Bahari}, \citenamefont {Fainman},\ and\ \citenamefont
  {Kant{\'e}}}]{kodigala2017lasing}%
  \BibitemOpen
  \bibfield  {author} {\bibinfo {author} {\bibfnamefont {A.}~\bibnamefont
  {Kodigala}}, \bibinfo {author} {\bibfnamefont {T.}~\bibnamefont {Lepetit}},
  \bibinfo {author} {\bibfnamefont {Q.}~\bibnamefont {Gu}}, \bibinfo {author}
  {\bibfnamefont {B.}~\bibnamefont {Bahari}}, \bibinfo {author} {\bibfnamefont
  {Y.}~\bibnamefont {Fainman}}, \ and\ \bibinfo {author} {\bibfnamefont
  {B.}~\bibnamefont {Kant{\'e}}},\ }\href {\doibase 10/f9kvfp} {\bibfield
  {journal} {\bibinfo  {journal} {Nature}\ }\textbf {\bibinfo {volume} {541}},\
  \bibinfo {pages} {196} (\bibinfo {year} {2017})}\BibitemShut {NoStop}%
\bibitem [{\citenamefont {Wu}\ \emph {et~al.}(2020)\citenamefont {Wu},
  \citenamefont {Ha}, \citenamefont {Shendre}, \citenamefont {Durmusoglu},
  \citenamefont {Koh}, \citenamefont {Abujetas}, \citenamefont
  {S{\'a}nchez-Gil}, \citenamefont {Paniagua-Dominguez}, \citenamefont
  {Demir},\ and\ \citenamefont {Kuznetsov}}]{wu2020room}%
  \BibitemOpen
  \bibfield  {author} {\bibinfo {author} {\bibfnamefont {M.}~\bibnamefont
  {Wu}}, \bibinfo {author} {\bibfnamefont {S.~T.}\ \bibnamefont {Ha}}, \bibinfo
  {author} {\bibfnamefont {S.}~\bibnamefont {Shendre}}, \bibinfo {author}
  {\bibfnamefont {E.~G.}\ \bibnamefont {Durmusoglu}}, \bibinfo {author}
  {\bibfnamefont {W.-K.}\ \bibnamefont {Koh}}, \bibinfo {author} {\bibfnamefont
  {D.~R.}\ \bibnamefont {Abujetas}}, \bibinfo {author} {\bibfnamefont {J.~A.}\
  \bibnamefont {S{\'a}nchez-Gil}}, \bibinfo {author} {\bibfnamefont
  {R.}~\bibnamefont {Paniagua-Dominguez}}, \bibinfo {author} {\bibfnamefont
  {H.~V.}\ \bibnamefont {Demir}}, \ and\ \bibinfo {author} {\bibfnamefont
  {A.~I.}\ \bibnamefont {Kuznetsov}},\ }\href@noop {} {\bibfield  {journal}
  {\bibinfo  {journal} {Nano Lett.}\ }\textbf {\bibinfo {volume} {20}},\
  \bibinfo {pages} {6005} (\bibinfo {year} {2020})}\BibitemShut {NoStop}%
\bibitem [{\citenamefont {Yang}\ \emph {et~al.}(2020)\citenamefont {Yang},
  \citenamefont {Maksimov}, \citenamefont {Huang}, \citenamefont {Pankin},
  \citenamefont {Timofeev}, \citenamefont {Hong}, \citenamefont {Li},
  \citenamefont {Chen}, \citenamefont {Hsu}, \citenamefont {Liu} \emph
  {et~al.}}]{yang2020low}%
  \BibitemOpen
  \bibfield  {author} {\bibinfo {author} {\bibfnamefont {J.-H.}\ \bibnamefont
  {Yang}}, \bibinfo {author} {\bibfnamefont {D.~N.}\ \bibnamefont {Maksimov}},
  \bibinfo {author} {\bibfnamefont {Z.-T.}\ \bibnamefont {Huang}}, \bibinfo
  {author} {\bibfnamefont {P.~S.}\ \bibnamefont {Pankin}}, \bibinfo {author}
  {\bibfnamefont {I.~V.}\ \bibnamefont {Timofeev}}, \bibinfo {author}
  {\bibfnamefont {K.-B.}\ \bibnamefont {Hong}}, \bibinfo {author}
  {\bibfnamefont {H.}~\bibnamefont {Li}}, \bibinfo {author} {\bibfnamefont
  {J.-W.}\ \bibnamefont {Chen}}, \bibinfo {author} {\bibfnamefont {C.-Y.}\
  \bibnamefont {Hsu}}, \bibinfo {author} {\bibfnamefont {Y.-Y.}\ \bibnamefont
  {Liu}},  \emph {et~al.},\ }\href@noop {} {\bibfield  {journal} {\bibinfo
  {journal} {arXiv preprint arXiv:2007.03233}\ } (\bibinfo {year}
  {2020})}\BibitemShut {NoStop}%
\bibitem [{\citenamefont {Romano}\ \emph {et~al.}(2018)\citenamefont {Romano},
  \citenamefont {Zito}, \citenamefont {Torino}, \citenamefont {Calafiore},
  \citenamefont {Penzo}, \citenamefont {Coppola}, \citenamefont {Cabrini},
  \citenamefont {Rendina},\ and\ \citenamefont {Mocella}}]{romano2018label}%
  \BibitemOpen
  \bibfield  {author} {\bibinfo {author} {\bibfnamefont {S.}~\bibnamefont
  {Romano}}, \bibinfo {author} {\bibfnamefont {G.}~\bibnamefont {Zito}},
  \bibinfo {author} {\bibfnamefont {S.}~\bibnamefont {Torino}}, \bibinfo
  {author} {\bibfnamefont {G.}~\bibnamefont {Calafiore}}, \bibinfo {author}
  {\bibfnamefont {E.}~\bibnamefont {Penzo}}, \bibinfo {author} {\bibfnamefont
  {G.}~\bibnamefont {Coppola}}, \bibinfo {author} {\bibfnamefont
  {S.}~\bibnamefont {Cabrini}}, \bibinfo {author} {\bibfnamefont
  {I.}~\bibnamefont {Rendina}}, \ and\ \bibinfo {author} {\bibfnamefont
  {V.}~\bibnamefont {Mocella}},\ }\href@noop {} {\bibfield  {journal} {\bibinfo
   {journal} {Photonics Res.}\ }\textbf {\bibinfo {volume} {6}},\ \bibinfo
  {pages} {726} (\bibinfo {year} {2018})}\BibitemShut {NoStop}%
\bibitem [{\citenamefont {Tittl}\ \emph {et~al.}(2018)\citenamefont {Tittl},
  \citenamefont {Leitis}, \citenamefont {Liu}, \citenamefont {Yesilkoy},
  \citenamefont {Choi}, \citenamefont {Neshev}, \citenamefont {Kivshar},\ and\
  \citenamefont {Altug}}]{tittl2018imagingbased}%
  \BibitemOpen
  \bibfield  {author} {\bibinfo {author} {\bibfnamefont {A.}~\bibnamefont
  {Tittl}}, \bibinfo {author} {\bibfnamefont {A.}~\bibnamefont {Leitis}},
  \bibinfo {author} {\bibfnamefont {M.}~\bibnamefont {Liu}}, \bibinfo {author}
  {\bibfnamefont {F.}~\bibnamefont {Yesilkoy}}, \bibinfo {author}
  {\bibfnamefont {D.-Y.}\ \bibnamefont {Choi}}, \bibinfo {author}
  {\bibfnamefont {D.~N.}\ \bibnamefont {Neshev}}, \bibinfo {author}
  {\bibfnamefont {Y.~S.}\ \bibnamefont {Kivshar}}, \ and\ \bibinfo {author}
  {\bibfnamefont {H.}~\bibnamefont {Altug}},\ }\href {\doibase 10/gdmwj8}
  {\bibfield  {journal} {\bibinfo  {journal} {Science}\ }\textbf {\bibinfo
  {volume} {360}},\ \bibinfo {pages} {1105} (\bibinfo {year}
  {2018})}\BibitemShut {NoStop}%
\bibitem [{\citenamefont {Foley}\ \emph {et~al.}(2014)\citenamefont {Foley},
  \citenamefont {Young},\ and\ \citenamefont
  {Phillips}}]{foley2014symmetryprotected}%
  \BibitemOpen
  \bibfield  {author} {\bibinfo {author} {\bibfnamefont {J.~M.}\ \bibnamefont
  {Foley}}, \bibinfo {author} {\bibfnamefont {S.~M.}\ \bibnamefont {Young}}, \
  and\ \bibinfo {author} {\bibfnamefont {J.~D.}\ \bibnamefont {Phillips}},\
  }\href {\doibase 10/ghspf2} {\bibfield  {journal} {\bibinfo  {journal} {Phys.
  Rev. B}\ }\textbf {\bibinfo {volume} {89}},\ \bibinfo {pages} {165111}
  (\bibinfo {year} {2014})}\BibitemShut {NoStop}%
\bibitem [{\citenamefont {Doskolovich}\ \emph {et~al.}(2019)\citenamefont
  {Doskolovich}, \citenamefont {Bezus},\ and\ \citenamefont
  {Bykov}}]{doskolovich2019integrated}%
  \BibitemOpen
  \bibfield  {author} {\bibinfo {author} {\bibfnamefont {L.~L.}\ \bibnamefont
  {Doskolovich}}, \bibinfo {author} {\bibfnamefont {E.~A.}\ \bibnamefont
  {Bezus}}, \ and\ \bibinfo {author} {\bibfnamefont {D.~A.}\ \bibnamefont
  {Bykov}},\ }\href {\doibase 10/ghspcp} {\bibfield  {journal} {\bibinfo
  {journal} {Photon. Res., PRJ}\ }\textbf {\bibinfo {volume} {7}},\ \bibinfo
  {pages} {1314} (\bibinfo {year} {2019})}\BibitemShut {NoStop}%
\bibitem [{\citenamefont {Koshelev}\ \emph
  {et~al.}(2018{\natexlab{a}})\citenamefont {Koshelev}, \citenamefont {Sychev},
  \citenamefont {Sadrieva}, \citenamefont {Bogdanov},\ and\ \citenamefont
  {Iorsh}}]{koshelev2018strong}%
  \BibitemOpen
  \bibfield  {author} {\bibinfo {author} {\bibfnamefont {K.~L.}\ \bibnamefont
  {Koshelev}}, \bibinfo {author} {\bibfnamefont {S.~K.}\ \bibnamefont
  {Sychev}}, \bibinfo {author} {\bibfnamefont {Z.~F.}\ \bibnamefont
  {Sadrieva}}, \bibinfo {author} {\bibfnamefont {A.~A.}\ \bibnamefont
  {Bogdanov}}, \ and\ \bibinfo {author} {\bibfnamefont {I.~V.}\ \bibnamefont
  {Iorsh}},\ }\href {\doibase 10/gfs355} {\bibfield  {journal} {\bibinfo
  {journal} {Phys. Rev. B}\ }\textbf {\bibinfo {volume} {98}},\ \bibinfo
  {pages} {161113} (\bibinfo {year} {2018}{\natexlab{a}})}\BibitemShut
  {NoStop}%
\bibitem [{\citenamefont {Kravtsov}\ \emph {et~al.}(2020)\citenamefont
  {Kravtsov}, \citenamefont {Khestanova}, \citenamefont {Benimetskiy},
  \citenamefont {Ivanova}, \citenamefont {Samusev}, \citenamefont {Sinev},
  \citenamefont {Pidgayko}, \citenamefont {Mozharov}, \citenamefont {Mukhin},
  \citenamefont {Lozhkin} \emph {et~al.}}]{kravtsov2020nonlinear}%
  \BibitemOpen
  \bibfield  {author} {\bibinfo {author} {\bibfnamefont {V.}~\bibnamefont
  {Kravtsov}}, \bibinfo {author} {\bibfnamefont {E.}~\bibnamefont
  {Khestanova}}, \bibinfo {author} {\bibfnamefont {F.~A.}\ \bibnamefont
  {Benimetskiy}}, \bibinfo {author} {\bibfnamefont {T.}~\bibnamefont
  {Ivanova}}, \bibinfo {author} {\bibfnamefont {A.~K.}\ \bibnamefont
  {Samusev}}, \bibinfo {author} {\bibfnamefont {I.~S.}\ \bibnamefont {Sinev}},
  \bibinfo {author} {\bibfnamefont {D.}~\bibnamefont {Pidgayko}}, \bibinfo
  {author} {\bibfnamefont {A.~M.}\ \bibnamefont {Mozharov}}, \bibinfo {author}
  {\bibfnamefont {I.~S.}\ \bibnamefont {Mukhin}}, \bibinfo {author}
  {\bibfnamefont {M.~S.}\ \bibnamefont {Lozhkin}},  \emph {et~al.},\
  }\href@noop {} {\bibfield  {journal} {\bibinfo  {journal} {Light Sci. Appl.}\
  }\textbf {\bibinfo {volume} {9}},\ \bibinfo {pages} {1} (\bibinfo {year}
  {2020})}\BibitemShut {NoStop}%
\bibitem [{\citenamefont {Bulgakov}\ and\ \citenamefont
  {Sadreev}(2014)}]{bulgakov2014robust}%
  \BibitemOpen
  \bibfield  {author} {\bibinfo {author} {\bibfnamefont {E.}~\bibnamefont
  {Bulgakov}}\ and\ \bibinfo {author} {\bibfnamefont {A.}~\bibnamefont
  {Sadreev}},\ }\href@noop {} {\bibfield  {journal} {\bibinfo  {journal} {Opt.
  Lett.}\ }\textbf {\bibinfo {volume} {39}},\ \bibinfo {pages} {5212} (\bibinfo
  {year} {2014})}\BibitemShut {NoStop}%
\bibitem [{\citenamefont {Yuan}\ and\ \citenamefont
  {Lu}(2016)}]{yuan2016diffraction}%
  \BibitemOpen
  \bibfield  {author} {\bibinfo {author} {\bibfnamefont {L.}~\bibnamefont
  {Yuan}}\ and\ \bibinfo {author} {\bibfnamefont {Y.~Y.}\ \bibnamefont {Lu}},\
  }\href {\doibase 10/ggqpsh} {\bibfield  {journal} {\bibinfo  {journal} {Phys.
  Rev. A}\ }\textbf {\bibinfo {volume} {94}},\ \bibinfo {pages} {013852}
  (\bibinfo {year} {2016})}\BibitemShut {NoStop}%
\bibitem [{\citenamefont {Yuan}\ and\ \citenamefont
  {Lu}(2017)}]{yuan2017strong}%
  \BibitemOpen
  \bibfield  {author} {\bibinfo {author} {\bibfnamefont {L.}~\bibnamefont
  {Yuan}}\ and\ \bibinfo {author} {\bibfnamefont {Y.~Y.}\ \bibnamefont {Lu}},\
  }\href {\doibase 10/ghspff} {\bibfield  {journal} {\bibinfo  {journal} {Phys.
  Rev. A}\ }\textbf {\bibinfo {volume} {95}},\ \bibinfo {pages} {023834}
  (\bibinfo {year} {2017})}\BibitemShut {NoStop}%
\bibitem [{\citenamefont {Carletti}\ \emph {et~al.}(2018)\citenamefont
  {Carletti}, \citenamefont {Koshelev}, \citenamefont {De~Angelis},\ and\
  \citenamefont {Kivshar}}]{carletti2018giant}%
  \BibitemOpen
  \bibfield  {author} {\bibinfo {author} {\bibfnamefont {L.}~\bibnamefont
  {Carletti}}, \bibinfo {author} {\bibfnamefont {K.}~\bibnamefont {Koshelev}},
  \bibinfo {author} {\bibfnamefont {C.}~\bibnamefont {De~Angelis}}, \ and\
  \bibinfo {author} {\bibfnamefont {Y.}~\bibnamefont {Kivshar}},\ }\href
  {\doibase 10/gfxm7x} {\bibfield  {journal} {\bibinfo  {journal} {Phys. Rev.
  Lett.}\ }\textbf {\bibinfo {volume} {121}},\ \bibinfo {pages} {033903}
  (\bibinfo {year} {2018})}\BibitemShut {NoStop}%
\bibitem [{\citenamefont {Bulgakov}\ and\ \citenamefont
  {Maksimov}(2019)}]{bulgakov2019nonlinear}%
  \BibitemOpen
  \bibfield  {author} {\bibinfo {author} {\bibfnamefont {E.~N.}\ \bibnamefont
  {Bulgakov}}\ and\ \bibinfo {author} {\bibfnamefont {D.~N.}\ \bibnamefont
  {Maksimov}},\ }\href {\doibase 10.1038/natrevmats.2016.48} {\bibfield
  {journal} {\bibinfo  {journal} {Sci. Rep.}\ }\textbf {\bibinfo {volume}
  {9}},\ \bibinfo {pages} {1} (\bibinfo {year} {2019})}\BibitemShut {NoStop}%
\bibitem [{\citenamefont {Koshelev}\ \emph
  {et~al.}(2019{\natexlab{b}})\citenamefont {Koshelev}, \citenamefont
  {Bogdanov},\ and\ \citenamefont {Kivshar}}]{koshelev2019meta}%
  \BibitemOpen
  \bibfield  {author} {\bibinfo {author} {\bibfnamefont {K.}~\bibnamefont
  {Koshelev}}, \bibinfo {author} {\bibfnamefont {A.}~\bibnamefont {Bogdanov}},
  \ and\ \bibinfo {author} {\bibfnamefont {Y.}~\bibnamefont {Kivshar}},\
  }\href@noop {} {\bibfield  {journal} {\bibinfo  {journal} {Sci. Bull.}\
  }\textbf {\bibinfo {volume} {64}},\ \bibinfo {pages} {836} (\bibinfo {year}
  {2019}{\natexlab{b}})}\BibitemShut {NoStop}%
\bibitem [{\citenamefont {Koshelev}\ \emph {et~al.}(2020)\citenamefont
  {Koshelev}, \citenamefont {Kruk}, \citenamefont {{Melik-Gaykazyan}},
  \citenamefont {Choi}, \citenamefont {Bogdanov}, \citenamefont {Park},\ and\
  \citenamefont {Kivshar}}]{koshelev2020subwavelength}%
  \BibitemOpen
  \bibfield  {author} {\bibinfo {author} {\bibfnamefont {K.}~\bibnamefont
  {Koshelev}}, \bibinfo {author} {\bibfnamefont {S.}~\bibnamefont {Kruk}},
  \bibinfo {author} {\bibfnamefont {E.}~\bibnamefont {{Melik-Gaykazyan}}},
  \bibinfo {author} {\bibfnamefont {J.-H.}\ \bibnamefont {Choi}}, \bibinfo
  {author} {\bibfnamefont {A.}~\bibnamefont {Bogdanov}}, \bibinfo {author}
  {\bibfnamefont {H.-G.}\ \bibnamefont {Park}}, \ and\ \bibinfo {author}
  {\bibfnamefont {Y.}~\bibnamefont {Kivshar}},\ }\href {\doibase 10/ghfr93}
  {\bibfield  {journal} {\bibinfo  {journal} {Science}\ }\textbf {\bibinfo
  {volume} {367}},\ \bibinfo {pages} {288} (\bibinfo {year}
  {2020})}\BibitemShut {NoStop}%
\bibitem [{\citenamefont {Minkov}\ \emph {et~al.}(2019)\citenamefont {Minkov},
  \citenamefont {Gerace},\ and\ \citenamefont {Fan}}]{minkov2019doubly}%
  \BibitemOpen
  \bibfield  {author} {\bibinfo {author} {\bibfnamefont {M.}~\bibnamefont
  {Minkov}}, \bibinfo {author} {\bibfnamefont {D.}~\bibnamefont {Gerace}}, \
  and\ \bibinfo {author} {\bibfnamefont {S.}~\bibnamefont {Fan}},\ }\href@noop
  {} {\bibfield  {journal} {\bibinfo  {journal} {Optica}\ }\textbf {\bibinfo
  {volume} {6}},\ \bibinfo {pages} {1039} (\bibinfo {year} {2019})}\BibitemShut
  {NoStop}%
\bibitem [{\citenamefont {Maksimov}\ \emph {et~al.}(2020)\citenamefont
  {Maksimov}, \citenamefont {Bogdanov},\ and\ \citenamefont
  {Bulgakov}}]{maksimov2020optical}%
  \BibitemOpen
  \bibfield  {author} {\bibinfo {author} {\bibfnamefont {D.~N.}\ \bibnamefont
  {Maksimov}}, \bibinfo {author} {\bibfnamefont {A.~A.}\ \bibnamefont
  {Bogdanov}}, \ and\ \bibinfo {author} {\bibfnamefont {E.~N.}\ \bibnamefont
  {Bulgakov}},\ }\href {\doibase 10/ghspdf} {\bibfield  {journal} {\bibinfo
  {journal} {Phys. Rev. A}\ }\textbf {\bibinfo {volume} {102}},\ \bibinfo
  {pages} {033511} (\bibinfo {year} {2020})}\BibitemShut {NoStop}%
\bibitem [{\citenamefont {Carletti}\ \emph {et~al.}(2019)\citenamefont
  {Carletti}, \citenamefont {Kruk}, \citenamefont {Bogdanov}, \citenamefont
  {De~Angelis},\ and\ \citenamefont {Kivshar}}]{carletti2019high}%
  \BibitemOpen
  \bibfield  {author} {\bibinfo {author} {\bibfnamefont {L.}~\bibnamefont
  {Carletti}}, \bibinfo {author} {\bibfnamefont {S.~S.}\ \bibnamefont {Kruk}},
  \bibinfo {author} {\bibfnamefont {A.~A.}\ \bibnamefont {Bogdanov}}, \bibinfo
  {author} {\bibfnamefont {C.}~\bibnamefont {De~Angelis}}, \ and\ \bibinfo
  {author} {\bibfnamefont {Y.}~\bibnamefont {Kivshar}},\ }\href@noop {}
  {\bibfield  {journal} {\bibinfo  {journal} {Phys. Rev. Res.}\ }\textbf
  {\bibinfo {volume} {1}},\ \bibinfo {pages} {023016} (\bibinfo {year}
  {2019})}\BibitemShut {NoStop}%
\bibitem [{\citenamefont {Doeleman}\ \emph {et~al.}(2018)\citenamefont
  {Doeleman}, \citenamefont {Monticone}, \citenamefont {den Hollander},
  \citenamefont {Al{\`u}},\ and\ \citenamefont
  {Koenderink}}]{doeleman2018experimental}%
  \BibitemOpen
  \bibfield  {author} {\bibinfo {author} {\bibfnamefont {H.~M.}\ \bibnamefont
  {Doeleman}}, \bibinfo {author} {\bibfnamefont {F.}~\bibnamefont {Monticone}},
  \bibinfo {author} {\bibfnamefont {W.}~\bibnamefont {den Hollander}}, \bibinfo
  {author} {\bibfnamefont {A.}~\bibnamefont {Al{\`u}}}, \ and\ \bibinfo
  {author} {\bibfnamefont {A.~F.}\ \bibnamefont {Koenderink}},\ }\href@noop {}
  {\bibfield  {journal} {\bibinfo  {journal} {Nat. Photonics}\ }\textbf
  {\bibinfo {volume} {12}},\ \bibinfo {pages} {397} (\bibinfo {year}
  {2018})}\BibitemShut {NoStop}%
\bibitem [{\citenamefont {Wang}\ \emph {et~al.}(2020)\citenamefont {Wang},
  \citenamefont {Liu}, \citenamefont {Zhao}, \citenamefont {Wang},
  \citenamefont {Zhang}, \citenamefont {Chen}, \citenamefont {Guan},
  \citenamefont {Liu}, \citenamefont {Shi},\ and\ \citenamefont
  {Zi}}]{wang2020generating}%
  \BibitemOpen
  \bibfield  {author} {\bibinfo {author} {\bibfnamefont {B.}~\bibnamefont
  {Wang}}, \bibinfo {author} {\bibfnamefont {W.}~\bibnamefont {Liu}}, \bibinfo
  {author} {\bibfnamefont {M.}~\bibnamefont {Zhao}}, \bibinfo {author}
  {\bibfnamefont {J.}~\bibnamefont {Wang}}, \bibinfo {author} {\bibfnamefont
  {Y.}~\bibnamefont {Zhang}}, \bibinfo {author} {\bibfnamefont
  {A.}~\bibnamefont {Chen}}, \bibinfo {author} {\bibfnamefont {F.}~\bibnamefont
  {Guan}}, \bibinfo {author} {\bibfnamefont {X.}~\bibnamefont {Liu}}, \bibinfo
  {author} {\bibfnamefont {L.}~\bibnamefont {Shi}}, \ and\ \bibinfo {author}
  {\bibfnamefont {J.}~\bibnamefont {Zi}},\ }\href {\doibase 10/gg34dp}
  {\bibfield  {journal} {\bibinfo  {journal} {Nat. Photonics}\ }\textbf
  {\bibinfo {volume} {14}},\ \bibinfo {pages} {623} (\bibinfo {year}
  {2020})}\BibitemShut {NoStop}%
\bibitem [{\citenamefont {Friedrich}\ and\ \citenamefont
  {Wintgen}(1985)}]{friedrich1985interfering}%
  \BibitemOpen
  \bibfield  {author} {\bibinfo {author} {\bibfnamefont {H.}~\bibnamefont
  {Friedrich}}\ and\ \bibinfo {author} {\bibfnamefont {D.}~\bibnamefont
  {Wintgen}},\ }\href {\doibase 10/ftvmw3} {\bibfield  {journal} {\bibinfo
  {journal} {Phys. Rev. A}\ }\textbf {\bibinfo {volume} {32}},\ \bibinfo
  {pages} {3231} (\bibinfo {year} {1985})}\BibitemShut {NoStop}%
\bibitem [{\citenamefont {Hsu}\ \emph {et~al.}(2013)\citenamefont {Hsu},
  \citenamefont {Zhen}, \citenamefont {Chua}, \citenamefont {Johnson},
  \citenamefont {Joannopoulos},\ and\ \citenamefont
  {Solja{\v{c}}i{\'c}}}]{hsu2013bloch}%
  \BibitemOpen
  \bibfield  {author} {\bibinfo {author} {\bibfnamefont {C.~W.}\ \bibnamefont
  {Hsu}}, \bibinfo {author} {\bibfnamefont {B.}~\bibnamefont {Zhen}}, \bibinfo
  {author} {\bibfnamefont {S.-L.}\ \bibnamefont {Chua}}, \bibinfo {author}
  {\bibfnamefont {S.~G.}\ \bibnamefont {Johnson}}, \bibinfo {author}
  {\bibfnamefont {J.~D.}\ \bibnamefont {Joannopoulos}}, \ and\ \bibinfo
  {author} {\bibfnamefont {M.}~\bibnamefont {Solja{\v{c}}i{\'c}}},\ }\href@noop
  {} {\bibfield  {journal} {\bibinfo  {journal} {Light Sci. Appl.}\ }\textbf
  {\bibinfo {volume} {2}},\ \bibinfo {pages} {e84} (\bibinfo {year}
  {2013})}\BibitemShut {NoStop}%
\bibitem [{\citenamefont {Zhen}\ \emph {et~al.}(2014)\citenamefont {Zhen},
  \citenamefont {Hsu}, \citenamefont {Lu}, \citenamefont {Stone},\ and\
  \citenamefont {Solja{\v{c}}i{\'c}}}]{zhen2014topological}%
  \BibitemOpen
  \bibfield  {author} {\bibinfo {author} {\bibfnamefont {B.}~\bibnamefont
  {Zhen}}, \bibinfo {author} {\bibfnamefont {C.~W.}\ \bibnamefont {Hsu}},
  \bibinfo {author} {\bibfnamefont {L.}~\bibnamefont {Lu}}, \bibinfo {author}
  {\bibfnamefont {A.~D.}\ \bibnamefont {Stone}}, \ and\ \bibinfo {author}
  {\bibfnamefont {M.}~\bibnamefont {Solja{\v{c}}i{\'c}}},\ }\href@noop {}
  {\bibfield  {journal} {\bibinfo  {journal} {Phys. Review Lett.}\ }\textbf
  {\bibinfo {volume} {113}},\ \bibinfo {pages} {257401} (\bibinfo {year}
  {2014})}\BibitemShut {NoStop}%
\bibitem [{\citenamefont {Sadrieva}\ \emph {et~al.}(2019)\citenamefont
  {Sadrieva}, \citenamefont {Frizyuk}, \citenamefont {Petrov}, \citenamefont
  {Kivshar},\ and\ \citenamefont {Bogdanov}}]{sadrieva2019multipolar}%
  \BibitemOpen
  \bibfield  {author} {\bibinfo {author} {\bibfnamefont {Z.}~\bibnamefont
  {Sadrieva}}, \bibinfo {author} {\bibfnamefont {K.}~\bibnamefont {Frizyuk}},
  \bibinfo {author} {\bibfnamefont {M.}~\bibnamefont {Petrov}}, \bibinfo
  {author} {\bibfnamefont {Y.}~\bibnamefont {Kivshar}}, \ and\ \bibinfo
  {author} {\bibfnamefont {A.}~\bibnamefont {Bogdanov}},\ }\href@noop {}
  {\bibfield  {journal} {\bibinfo  {journal} {Phys. Rev. B}\ }\textbf {\bibinfo
  {volume} {100}},\ \bibinfo {pages} {115303} (\bibinfo {year}
  {2019})}\BibitemShut {NoStop}%
\bibitem [{\citenamefont {Overvig}\ \emph {et~al.}(2020)\citenamefont
  {Overvig}, \citenamefont {Malek}, \citenamefont {Carter}, \citenamefont
  {Shrestha},\ and\ \citenamefont {Yu}}]{Overvig2020Selection}%
  \BibitemOpen
  \bibfield  {author} {\bibinfo {author} {\bibfnamefont {A.~C.}\ \bibnamefont
  {Overvig}}, \bibinfo {author} {\bibfnamefont {S.~C.}\ \bibnamefont {Malek}},
  \bibinfo {author} {\bibfnamefont {M.~J.}\ \bibnamefont {Carter}}, \bibinfo
  {author} {\bibfnamefont {S.}~\bibnamefont {Shrestha}}, \ and\ \bibinfo
  {author} {\bibfnamefont {N.}~\bibnamefont {Yu}},\ }\href {\doibase
  10.1103/PhysRevB.102.035434} {\bibfield  {journal} {\bibinfo  {journal}
  {Phys. Rev. B}\ }\textbf {\bibinfo {volume} {102}},\ \bibinfo {pages}
  {035434} (\bibinfo {year} {2020})}\BibitemShut {NoStop}%
\bibitem [{\citenamefont {Wu}\ \emph {et~al.}(2012)\citenamefont {Wu},
  \citenamefont {Khanikaev}, \citenamefont {Adato}, \citenamefont {Arju},
  \citenamefont {Yanik}, \citenamefont {Altug},\ and\ \citenamefont
  {Shvets}}]{wu2012fano}%
  \BibitemOpen
  \bibfield  {author} {\bibinfo {author} {\bibfnamefont {C.}~\bibnamefont
  {Wu}}, \bibinfo {author} {\bibfnamefont {A.~B.}\ \bibnamefont {Khanikaev}},
  \bibinfo {author} {\bibfnamefont {R.}~\bibnamefont {Adato}}, \bibinfo
  {author} {\bibfnamefont {N.}~\bibnamefont {Arju}}, \bibinfo {author}
  {\bibfnamefont {A.~A.}\ \bibnamefont {Yanik}}, \bibinfo {author}
  {\bibfnamefont {H.}~\bibnamefont {Altug}}, \ and\ \bibinfo {author}
  {\bibfnamefont {G.}~\bibnamefont {Shvets}},\ }\href@noop {} {\bibfield
  {journal} {\bibinfo  {journal} {Nat. Mater.}\ }\textbf {\bibinfo {volume}
  {11}},\ \bibinfo {pages} {69} (\bibinfo {year} {2012})}\BibitemShut {NoStop}%
\bibitem [{\citenamefont {Mouadili}\ \emph {et~al.}(2013)\citenamefont
  {Mouadili}, \citenamefont {El~Boudouti}, \citenamefont {Soltani},
  \citenamefont {Talbi}, \citenamefont {Akjouj},\ and\ \citenamefont
  {Djafari-Rouhani}}]{mouadili2013theoretical}%
  \BibitemOpen
  \bibfield  {author} {\bibinfo {author} {\bibfnamefont {A.}~\bibnamefont
  {Mouadili}}, \bibinfo {author} {\bibfnamefont {E.}~\bibnamefont
  {El~Boudouti}}, \bibinfo {author} {\bibfnamefont {A.}~\bibnamefont
  {Soltani}}, \bibinfo {author} {\bibfnamefont {A.}~\bibnamefont {Talbi}},
  \bibinfo {author} {\bibfnamefont {A.}~\bibnamefont {Akjouj}}, \ and\ \bibinfo
  {author} {\bibfnamefont {B.}~\bibnamefont {Djafari-Rouhani}},\ }\href@noop {}
  {\bibfield  {journal} {\bibinfo  {journal} {Journal of Applied Physics}\
  }\textbf {\bibinfo {volume} {113}},\ \bibinfo {pages} {164101} (\bibinfo
  {year} {2013})}\BibitemShut {NoStop}%
\bibitem [{\citenamefont {Yanik}\ \emph {et~al.}(2011)\citenamefont {Yanik},
  \citenamefont {Cetin}, \citenamefont {Huang}, \citenamefont {Artar},
  \citenamefont {Mousavi}, \citenamefont {Khanikaev}, \citenamefont {Connor},
  \citenamefont {Shvets},\ and\ \citenamefont {Altug}}]{yanik2011seeing}%
  \BibitemOpen
  \bibfield  {author} {\bibinfo {author} {\bibfnamefont {A.~A.}\ \bibnamefont
  {Yanik}}, \bibinfo {author} {\bibfnamefont {A.~E.}\ \bibnamefont {Cetin}},
  \bibinfo {author} {\bibfnamefont {M.}~\bibnamefont {Huang}}, \bibinfo
  {author} {\bibfnamefont {A.}~\bibnamefont {Artar}}, \bibinfo {author}
  {\bibfnamefont {S.~H.}\ \bibnamefont {Mousavi}}, \bibinfo {author}
  {\bibfnamefont {A.}~\bibnamefont {Khanikaev}}, \bibinfo {author}
  {\bibfnamefont {J.~H.}\ \bibnamefont {Connor}}, \bibinfo {author}
  {\bibfnamefont {G.}~\bibnamefont {Shvets}}, \ and\ \bibinfo {author}
  {\bibfnamefont {H.}~\bibnamefont {Altug}},\ }\href {\doibase 10/fc3fgw}
  {\bibfield  {journal} {\bibinfo  {journal} {Proc. Natl. Acad. Sci.}\ }\textbf
  {\bibinfo {volume} {108}},\ \bibinfo {pages} {11784} (\bibinfo {year}
  {2011})}\BibitemShut {NoStop}%
\bibitem [{\citenamefont {Maier}(2006)}]{maier2006plasmonic}%
  \BibitemOpen
  \bibfield  {author} {\bibinfo {author} {\bibfnamefont {S.~A.}\ \bibnamefont
  {Maier}},\ }\href@noop {} {\bibfield  {journal} {\bibinfo  {journal} {Opt.
  Express}\ }\textbf {\bibinfo {volume} {14}},\ \bibinfo {pages} {1957}
  (\bibinfo {year} {2006})}\BibitemShut {NoStop}%
\bibitem [{\citenamefont {Seok}\ \emph {et~al.}(2011)\citenamefont {Seok},
  \citenamefont {Jamshidi}, \citenamefont {Kim}, \citenamefont {Dhuey},
  \citenamefont {Lakhani}, \citenamefont {Choo}, \citenamefont {Schuck},
  \citenamefont {Cabrini}, \citenamefont {Schwartzberg}, \citenamefont {Bokor}
  \emph {et~al.}}]{seok2011radiation}%
  \BibitemOpen
  \bibfield  {author} {\bibinfo {author} {\bibfnamefont {T.~J.}\ \bibnamefont
  {Seok}}, \bibinfo {author} {\bibfnamefont {A.}~\bibnamefont {Jamshidi}},
  \bibinfo {author} {\bibfnamefont {M.}~\bibnamefont {Kim}}, \bibinfo {author}
  {\bibfnamefont {S.}~\bibnamefont {Dhuey}}, \bibinfo {author} {\bibfnamefont
  {A.}~\bibnamefont {Lakhani}}, \bibinfo {author} {\bibfnamefont
  {H.}~\bibnamefont {Choo}}, \bibinfo {author} {\bibfnamefont {P.~J.}\
  \bibnamefont {Schuck}}, \bibinfo {author} {\bibfnamefont {S.}~\bibnamefont
  {Cabrini}}, \bibinfo {author} {\bibfnamefont {A.~M.}\ \bibnamefont
  {Schwartzberg}}, \bibinfo {author} {\bibfnamefont {J.}~\bibnamefont {Bokor}},
   \emph {et~al.},\ }\href@noop {} {\bibfield  {journal} {\bibinfo  {journal}
  {Nano Lett.}\ }\textbf {\bibinfo {volume} {11}},\ \bibinfo {pages} {2606}
  (\bibinfo {year} {2011})}\BibitemShut {NoStop}%
\bibitem [{\citenamefont {Dyakov}\ \emph {et~al.}(2020)\citenamefont {Dyakov},
  \citenamefont {Stepikhova}, \citenamefont {Bogdanov}, \citenamefont
  {Novikov}, \citenamefont {Yurasov}, \citenamefont {Krasilnik}, \citenamefont
  {Tikhodeev},\ and\ \citenamefont {Gippius}}]{dyakov2020photonic}%
  \BibitemOpen
  \bibfield  {author} {\bibinfo {author} {\bibfnamefont {S.~A.}\ \bibnamefont
  {Dyakov}}, \bibinfo {author} {\bibfnamefont {M.~V.}\ \bibnamefont
  {Stepikhova}}, \bibinfo {author} {\bibfnamefont {A.~A.}\ \bibnamefont
  {Bogdanov}}, \bibinfo {author} {\bibfnamefont {A.~V.}\ \bibnamefont
  {Novikov}}, \bibinfo {author} {\bibfnamefont {D.~V.}\ \bibnamefont
  {Yurasov}}, \bibinfo {author} {\bibfnamefont {Z.~F.}\ \bibnamefont
  {Krasilnik}}, \bibinfo {author} {\bibfnamefont {S.~G.}\ \bibnamefont
  {Tikhodeev}}, \ and\ \bibinfo {author} {\bibfnamefont {N.~A.}\ \bibnamefont
  {Gippius}},\ }\href@noop {} {\bibfield  {journal} {\bibinfo  {journal} {arXiv
  preprint arXiv:2006.06086}\ } (\bibinfo {year} {2020})}\BibitemShut {NoStop}%
\bibitem [{\citenamefont {Zhu}\ \emph {et~al.}(2020)\citenamefont {Zhu},
  \citenamefont {Yuan}, \citenamefont {Zeng},\ and\ \citenamefont
  {Xia}}]{zhu2020manipulating}%
  \BibitemOpen
  \bibfield  {author} {\bibinfo {author} {\bibfnamefont {L.}~\bibnamefont
  {Zhu}}, \bibinfo {author} {\bibfnamefont {S.}~\bibnamefont {Yuan}}, \bibinfo
  {author} {\bibfnamefont {C.}~\bibnamefont {Zeng}}, \ and\ \bibinfo {author}
  {\bibfnamefont {J.}~\bibnamefont {Xia}},\ }\href@noop {} {\bibfield
  {journal} {\bibinfo  {journal} {Adv. Opt. Mater.}\ }\textbf {\bibinfo
  {volume} {8}},\ \bibinfo {pages} {1901830} (\bibinfo {year}
  {2020})}\BibitemShut {NoStop}%
\bibitem [{\citenamefont {Seo}\ \emph {et~al.}(2020)\citenamefont {Seo},
  \citenamefont {Kim}, \citenamefont {Woo}, \citenamefont {Chung},\ and\
  \citenamefont {Jun}}]{seo2020fourier}%
  \BibitemOpen
  \bibfield  {author} {\bibinfo {author} {\bibfnamefont {I.~C.}\ \bibnamefont
  {Seo}}, \bibinfo {author} {\bibfnamefont {S.}~\bibnamefont {Kim}}, \bibinfo
  {author} {\bibfnamefont {B.~H.}\ \bibnamefont {Woo}}, \bibinfo {author}
  {\bibfnamefont {I.-S.}\ \bibnamefont {Chung}}, \ and\ \bibinfo {author}
  {\bibfnamefont {Y.~C.}\ \bibnamefont {Jun}},\ }\href@noop {} {\bibfield
  {journal} {\bibinfo  {journal} {Nanophotonics}\ }\textbf {\bibinfo {volume}
  {9}},\ \bibinfo {pages} {4565} (\bibinfo {year} {2020})}\BibitemShut
  {NoStop}%
\bibitem [{\citenamefont {Koshelev}\ \emph
  {et~al.}(2018{\natexlab{b}})\citenamefont {Koshelev}, \citenamefont
  {Lepeshov}, \citenamefont {Liu}, \citenamefont {Bogdanov},\ and\
  \citenamefont {Kivshar}}]{koshelev2018asymmetric}%
  \BibitemOpen
  \bibfield  {author} {\bibinfo {author} {\bibfnamefont {K.}~\bibnamefont
  {Koshelev}}, \bibinfo {author} {\bibfnamefont {S.}~\bibnamefont {Lepeshov}},
  \bibinfo {author} {\bibfnamefont {M.}~\bibnamefont {Liu}}, \bibinfo {author}
  {\bibfnamefont {A.}~\bibnamefont {Bogdanov}}, \ and\ \bibinfo {author}
  {\bibfnamefont {Y.}~\bibnamefont {Kivshar}},\ }\href@noop {} {\bibfield
  {journal} {\bibinfo  {journal} {Phys. Rev. Lett.}\ }\textbf {\bibinfo
  {volume} {121}},\ \bibinfo {pages} {193903} (\bibinfo {year}
  {2018}{\natexlab{b}})}\BibitemShut {NoStop}%
\bibitem [{\citenamefont {Deriy}\ \emph {et~al.}(2021)\citenamefont {Deriy},
  \citenamefont {Toftul}, \citenamefont {Petrov},\ and\ \citenamefont
  {Bogdanov}}]{deriy2021bound}%
  \BibitemOpen
  \bibfield  {author} {\bibinfo {author} {\bibfnamefont {I.}~\bibnamefont
  {Deriy}}, \bibinfo {author} {\bibfnamefont {I.}~\bibnamefont {Toftul}},
  \bibinfo {author} {\bibfnamefont {M.}~\bibnamefont {Petrov}}, \ and\ \bibinfo
  {author} {\bibfnamefont {A.}~\bibnamefont {Bogdanov}},\ }\href@noop {}
  {\bibfield  {journal} {\bibinfo  {journal} {arXiv preprint arXiv:2104.05539}\
  } (\bibinfo {year} {2021})}\BibitemShut {NoStop}%
\bibitem [{\citenamefont {Bernhardt}\ \emph {et~al.}(2020)\citenamefont
  {Bernhardt}, \citenamefont {Koshelev}, \citenamefont {White}, \citenamefont
  {Meng}, \citenamefont {Froch}, \citenamefont {Kim}, \citenamefont {Tran},
  \citenamefont {Choi}, \citenamefont {Kivshar},\ and\ \citenamefont
  {Solntsev}}]{bernhardt2020quasi}%
  \BibitemOpen
  \bibfield  {author} {\bibinfo {author} {\bibfnamefont {N.}~\bibnamefont
  {Bernhardt}}, \bibinfo {author} {\bibfnamefont {K.}~\bibnamefont {Koshelev}},
  \bibinfo {author} {\bibfnamefont {S.~J.}\ \bibnamefont {White}}, \bibinfo
  {author} {\bibfnamefont {K.~W.~C.}\ \bibnamefont {Meng}}, \bibinfo {author}
  {\bibfnamefont {J.~E.}\ \bibnamefont {Froch}}, \bibinfo {author}
  {\bibfnamefont {S.}~\bibnamefont {Kim}}, \bibinfo {author} {\bibfnamefont
  {T.~T.}\ \bibnamefont {Tran}}, \bibinfo {author} {\bibfnamefont {D.-Y.}\
  \bibnamefont {Choi}}, \bibinfo {author} {\bibfnamefont {Y.}~\bibnamefont
  {Kivshar}}, \ and\ \bibinfo {author} {\bibfnamefont {A.~S.}\ \bibnamefont
  {Solntsev}},\ }\href@noop {} {\bibfield  {journal} {\bibinfo  {journal} {Nano
  Lett.}\ }\textbf {\bibinfo {volume} {20}},\ \bibinfo {pages} {5309} (\bibinfo
  {year} {2020})}\BibitemShut {NoStop}%
\bibitem [{\citenamefont {Zograf}\ \emph {et~al.}(2020)\citenamefont {Zograf},
  \citenamefont {Koshelev}, \citenamefont {Zalogina}, \citenamefont {Korolev},
  \citenamefont {Choi}, \citenamefont {Zurch}, \citenamefont {Spielmann},
  \citenamefont {Luther-Davies}, \citenamefont {Kartashov}, \citenamefont
  {Makarov} \emph {et~al.}}]{zograf2020high}%
  \BibitemOpen
  \bibfield  {author} {\bibinfo {author} {\bibfnamefont {G.}~\bibnamefont
  {Zograf}}, \bibinfo {author} {\bibfnamefont {K.}~\bibnamefont {Koshelev}},
  \bibinfo {author} {\bibfnamefont {A.}~\bibnamefont {Zalogina}}, \bibinfo
  {author} {\bibfnamefont {V.}~\bibnamefont {Korolev}}, \bibinfo {author}
  {\bibfnamefont {D.-Y.}\ \bibnamefont {Choi}}, \bibinfo {author}
  {\bibfnamefont {M.}~\bibnamefont {Zurch}}, \bibinfo {author} {\bibfnamefont
  {C.}~\bibnamefont {Spielmann}}, \bibinfo {author} {\bibfnamefont
  {B.}~\bibnamefont {Luther-Davies}}, \bibinfo {author} {\bibfnamefont
  {D.}~\bibnamefont {Kartashov}}, \bibinfo {author} {\bibfnamefont
  {S.}~\bibnamefont {Makarov}},  \emph {et~al.},\ }\href@noop {} {\bibfield
  {journal} {\bibinfo  {journal} {arXiv preprint arXiv:2008.11481}\ } (\bibinfo
  {year} {2020})}\BibitemShut {NoStop}%
\bibitem [{\citenamefont {Anthur}\ \emph {et~al.}(2020)\citenamefont {Anthur},
  \citenamefont {Zhang}, \citenamefont {{Paniagua-Dominguez}}, \citenamefont
  {Kalashnikov}, \citenamefont {Ha}, \citenamefont {Ma{\ss}}, \citenamefont
  {Kuznetsov},\ and\ \citenamefont {Krivitsky}}]{anthur2020continuous}%
  \BibitemOpen
  \bibfield  {author} {\bibinfo {author} {\bibfnamefont {A.~P.}\ \bibnamefont
  {Anthur}}, \bibinfo {author} {\bibfnamefont {H.}~\bibnamefont {Zhang}},
  \bibinfo {author} {\bibfnamefont {R.}~\bibnamefont {{Paniagua-Dominguez}}},
  \bibinfo {author} {\bibfnamefont {D.~A.}\ \bibnamefont {Kalashnikov}},
  \bibinfo {author} {\bibfnamefont {S.~T.}\ \bibnamefont {Ha}}, \bibinfo
  {author} {\bibfnamefont {T.~W.~W.}\ \bibnamefont {Ma{\ss}}}, \bibinfo
  {author} {\bibfnamefont {A.~I.}\ \bibnamefont {Kuznetsov}}, \ and\ \bibinfo
  {author} {\bibfnamefont {L.}~\bibnamefont {Krivitsky}},\ }\href {\doibase
  10/ghsgfv} {\bibfield  {journal} {\bibinfo  {journal} {Nano Lett.}\ }\textbf
  {\bibinfo {volume} {20}},\ \bibinfo {pages} {8745} (\bibinfo {year}
  {2020})}\BibitemShut {NoStop}%
\bibitem [{\citenamefont {Vyas}\ and\ \citenamefont
  {Hegde}(2020)}]{vyas2020improved}%
  \BibitemOpen
  \bibfield  {author} {\bibinfo {author} {\bibfnamefont {H.}~\bibnamefont
  {Vyas}}\ and\ \bibinfo {author} {\bibfnamefont {R.~S.}\ \bibnamefont
  {Hegde}},\ }\href@noop {} {\bibfield  {journal} {\bibinfo  {journal} {Opt.
  Mater. Express}\ }\textbf {\bibinfo {volume} {10}},\ \bibinfo {pages} {1616}
  (\bibinfo {year} {2020})}\BibitemShut {NoStop}%
\bibitem [{\citenamefont {Chen}\ \emph {et~al.}(2020)\citenamefont {Chen},
  \citenamefont {Zhao}, \citenamefont {Zhang},\ and\ \citenamefont
  {Qiu}}]{chen2020integrated}%
  \BibitemOpen
  \bibfield  {author} {\bibinfo {author} {\bibfnamefont {Y.}~\bibnamefont
  {Chen}}, \bibinfo {author} {\bibfnamefont {C.}~\bibnamefont {Zhao}}, \bibinfo
  {author} {\bibfnamefont {Y.}~\bibnamefont {Zhang}}, \ and\ \bibinfo {author}
  {\bibfnamefont {C.-w.}\ \bibnamefont {Qiu}},\ }\href {\doibase 10/ghsgfz}
  {\bibfield  {journal} {\bibinfo  {journal} {Nano Lett.}\ }\textbf {\bibinfo
  {volume} {20}},\ \bibinfo {pages} {8696} (\bibinfo {year}
  {2020})}\BibitemShut {NoStop}%
\bibitem [{\citenamefont {Tseng}\ \emph {et~al.}(2020)\citenamefont {Tseng},
  \citenamefont {Jahani}, \citenamefont {Leitis},\ and\ \citenamefont
  {Altug}}]{tseng2020dielectric}%
  \BibitemOpen
  \bibfield  {author} {\bibinfo {author} {\bibfnamefont {M.~L.}\ \bibnamefont
  {Tseng}}, \bibinfo {author} {\bibfnamefont {Y.}~\bibnamefont {Jahani}},
  \bibinfo {author} {\bibfnamefont {A.}~\bibnamefont {Leitis}}, \ and\ \bibinfo
  {author} {\bibfnamefont {H.}~\bibnamefont {Altug}},\ }\href@noop {}
  {\bibfield  {journal} {\bibinfo  {journal} {ACS Photonics}\ } (\bibinfo
  {year} {2020})}\BibitemShut {NoStop}%
\bibitem [{\citenamefont {Yuan}\ and\ \citenamefont
  {Lu}(2014)}]{yuan2014bilateral}%
  \BibitemOpen
  \bibfield  {author} {\bibinfo {author} {\bibfnamefont {L.}~\bibnamefont
  {Yuan}}\ and\ \bibinfo {author} {\bibfnamefont {Y.~Y.}\ \bibnamefont {Lu}},\
  }\href {\doibase 10/ghspfh} {\bibfield  {journal} {\bibinfo  {journal} {Opt.
  Express, OE}\ }\textbf {\bibinfo {volume} {22}},\ \bibinfo {pages} {30128}
  (\bibinfo {year} {2014})}\BibitemShut {NoStop}%
\bibitem [{\citenamefont {Bulgakov}\ \emph {et~al.}(2015)\citenamefont
  {Bulgakov}, \citenamefont {Pichugin},\ and\ \citenamefont
  {Sadreev}}]{bulgakov2015alloptical}%
  \BibitemOpen
  \bibfield  {author} {\bibinfo {author} {\bibfnamefont {E.~N.}\ \bibnamefont
  {Bulgakov}}, \bibinfo {author} {\bibfnamefont {K.~N.}\ \bibnamefont
  {Pichugin}}, \ and\ \bibinfo {author} {\bibfnamefont {A.~F.}\ \bibnamefont
  {Sadreev}},\ }\href {\doibase 10/ghmjqf} {\bibfield  {journal} {\bibinfo
  {journal} {Opt. Express}\ }\textbf {\bibinfo {volume} {23}},\ \bibinfo
  {pages} {22520} (\bibinfo {year} {2015})}\BibitemShut {NoStop}%
\bibitem [{\citenamefont {Krasikov}\ \emph {et~al.}(2018)\citenamefont
  {Krasikov}, \citenamefont {Bogdanov},\ and\ \citenamefont
  {Iorsh}}]{krasikov2018nonlinear}%
  \BibitemOpen
  \bibfield  {author} {\bibinfo {author} {\bibfnamefont {S.~D.}\ \bibnamefont
  {Krasikov}}, \bibinfo {author} {\bibfnamefont {A.~A.}\ \bibnamefont
  {Bogdanov}}, \ and\ \bibinfo {author} {\bibfnamefont {I.~V.}\ \bibnamefont
  {Iorsh}},\ }\href {\doibase 10/gg9fjq} {\bibfield  {journal} {\bibinfo
  {journal} {Phys. Rev. B}\ }\textbf {\bibinfo {volume} {97}},\ \bibinfo
  {pages} {224309} (\bibinfo {year} {2018})}\BibitemShut {NoStop}%
\bibitem [{\citenamefont {Volkovskaya}\ \emph {et~al.}(2020)\citenamefont
  {Volkovskaya}, \citenamefont {Xu}, \citenamefont {Huang}, \citenamefont
  {Smirnov}, \citenamefont {Miroshnichenko},\ and\ \citenamefont
  {Smirnova}}]{volkovskaya2020multipolar}%
  \BibitemOpen
  \bibfield  {author} {\bibinfo {author} {\bibfnamefont {I.}~\bibnamefont
  {Volkovskaya}}, \bibinfo {author} {\bibfnamefont {L.}~\bibnamefont {Xu}},
  \bibinfo {author} {\bibfnamefont {L.}~\bibnamefont {Huang}}, \bibinfo
  {author} {\bibfnamefont {A.~I.}\ \bibnamefont {Smirnov}}, \bibinfo {author}
  {\bibfnamefont {A.~E.}\ \bibnamefont {Miroshnichenko}}, \ and\ \bibinfo
  {author} {\bibfnamefont {D.}~\bibnamefont {Smirnova}},\ }\href {\doibase
  10/ghmjxn} {\bibfield  {journal} {\bibinfo  {journal} {Nanophotonics}\
  }\textbf {\bibinfo {volume} {9}},\ \bibinfo {pages} {3953} (\bibinfo {year}
  {2020})}\BibitemShut {NoStop}%
\bibitem [{\citenamefont {Yuan}\ and\ \citenamefont
  {Lu}(2020)}]{yuan2020excitation}%
  \BibitemOpen
  \bibfield  {author} {\bibinfo {author} {\bibfnamefont {L.}~\bibnamefont
  {Yuan}}\ and\ \bibinfo {author} {\bibfnamefont {Y.~Y.}\ \bibnamefont {Lu}},\
  }\href {\doibase 10/ghmjxm} {\bibfield  {journal} {\bibinfo  {journal} {SIAM
  J. Appl. Math.}\ }\textbf {\bibinfo {volume} {80}},\ \bibinfo {pages} {864}
  (\bibinfo {year} {2020})}\BibitemShut {NoStop}%
\bibitem [{\citenamefont {Lee}\ \emph {et~al.}(2020)\citenamefont {Lee},
  \citenamefont {Kim}, \citenamefont {Kee},\ and\ \citenamefont
  {Magnusson}}]{lee2020polarization}%
  \BibitemOpen
  \bibfield  {author} {\bibinfo {author} {\bibfnamefont {S.-G.}\ \bibnamefont
  {Lee}}, \bibinfo {author} {\bibfnamefont {S.-H.}\ \bibnamefont {Kim}},
  \bibinfo {author} {\bibfnamefont {C.-S.}\ \bibnamefont {Kee}}, \ and\
  \bibinfo {author} {\bibfnamefont {R.}~\bibnamefont {Magnusson}},\ }\href@noop
  {} {\bibfield  {journal} {\bibinfo  {journal} {Opt. Express}\ }\textbf
  {\bibinfo {volume} {28}},\ \bibinfo {pages} {39453} (\bibinfo {year}
  {2020})}\BibitemShut {NoStop}%
\bibitem [{\citenamefont {Malomed}(2013)}]{malomed2013spontaneous}%
  \BibitemOpen
  \bibfield  {author} {\bibinfo {author} {\bibfnamefont {B.~A.}\ \bibnamefont
  {Malomed}},\ }\href@noop {} {\emph {\bibinfo {title} {Spontaneous symmetry
  breaking, self-trapping, and Josephson oscillations}}}\ (\bibinfo
  {publisher} {Springer},\ \bibinfo {address} {New York},\ \bibinfo {year}
  {2013})\BibitemShut {NoStop}%
\bibitem [{\citenamefont {Mayteevarunyoo}\ \emph {et~al.}(2008)\citenamefont
  {Mayteevarunyoo}, \citenamefont {Malomed},\ and\ \citenamefont
  {Dong}}]{mayteevarunyoo2008spontaneous}%
  \BibitemOpen
  \bibfield  {author} {\bibinfo {author} {\bibfnamefont {T.}~\bibnamefont
  {Mayteevarunyoo}}, \bibinfo {author} {\bibfnamefont {B.~A.}\ \bibnamefont
  {Malomed}}, \ and\ \bibinfo {author} {\bibfnamefont {G.}~\bibnamefont
  {Dong}},\ }\href@noop {} {\bibfield  {journal} {\bibinfo  {journal} {Phys.
  Rev. A}\ }\textbf {\bibinfo {volume} {78}},\ \bibinfo {pages} {053601}
  (\bibinfo {year} {2008})}\BibitemShut {NoStop}%
\bibitem [{\citenamefont {Kevrekidis}\ \emph {et~al.}(2005)\citenamefont
  {Kevrekidis}, \citenamefont {Chen}, \citenamefont {Malomed}, \citenamefont
  {Frantzeskakis},\ and\ \citenamefont
  {Weinstein}}]{kevrekidis2005spontaneous}%
  \BibitemOpen
  \bibfield  {author} {\bibinfo {author} {\bibfnamefont {P.}~\bibnamefont
  {Kevrekidis}}, \bibinfo {author} {\bibfnamefont {Z.}~\bibnamefont {Chen}},
  \bibinfo {author} {\bibfnamefont {B.}~\bibnamefont {Malomed}}, \bibinfo
  {author} {\bibfnamefont {D.}~\bibnamefont {Frantzeskakis}}, \ and\ \bibinfo
  {author} {\bibfnamefont {M.}~\bibnamefont {Weinstein}},\ }\href@noop {}
  {\bibfield  {journal} {\bibinfo  {journal} {Phys. Lett. A}\ }\textbf
  {\bibinfo {volume} {340}},\ \bibinfo {pages} {275} (\bibinfo {year}
  {2005})}\BibitemShut {NoStop}%
\bibitem [{\citenamefont {Yariv}(1973)}]{yariv1973coupledmode}%
  \BibitemOpen
  \bibfield  {author} {\bibinfo {author} {\bibfnamefont {A.}~\bibnamefont
  {Yariv}},\ }\href {\doibase 10/fhcvhm} {\bibfield  {journal} {\bibinfo
  {journal} {IEEE J. Quantum Electron.}\ }\textbf {\bibinfo {volume} {9}},\
  \bibinfo {pages} {919} (\bibinfo {year} {1973})}\BibitemShut {NoStop}%
\bibitem [{\citenamefont {Haus}\ and\ \citenamefont
  {Huang}(1991)}]{haus1991coupledmode}%
  \BibitemOpen
  \bibfield  {author} {\bibinfo {author} {\bibfnamefont {H.~A.}\ \bibnamefont
  {Haus}}\ and\ \bibinfo {author} {\bibfnamefont {W.}~\bibnamefont {Huang}},\
  }\href {\doibase 10/b8xh66} {\bibfield  {journal} {\bibinfo  {journal} {Proc.
  IEEE}\ }\textbf {\bibinfo {volume} {79}},\ \bibinfo {pages} {1505} (\bibinfo
  {year} {1991})}\BibitemShut {NoStop}%
\bibitem [{\citenamefont {Ostrovskaya}\ \emph {et~al.}(2000)\citenamefont
  {Ostrovskaya}, \citenamefont {Kivshar}, \citenamefont {Lisak}, \citenamefont
  {Hall}, \citenamefont {Cattani},\ and\ \citenamefont
  {Anderson}}]{ostrovskaya2000coupledmode}%
  \BibitemOpen
  \bibfield  {author} {\bibinfo {author} {\bibfnamefont {E.~A.}\ \bibnamefont
  {Ostrovskaya}}, \bibinfo {author} {\bibfnamefont {Y.~S.}\ \bibnamefont
  {Kivshar}}, \bibinfo {author} {\bibfnamefont {M.}~\bibnamefont {Lisak}},
  \bibinfo {author} {\bibfnamefont {B.}~\bibnamefont {Hall}}, \bibinfo {author}
  {\bibfnamefont {F.}~\bibnamefont {Cattani}}, \ and\ \bibinfo {author}
  {\bibfnamefont {D.}~\bibnamefont {Anderson}},\ }\href {\doibase 10/d5qq43}
  {\bibfield  {journal} {\bibinfo  {journal} {Phys. Rev. A}\ }\textbf {\bibinfo
  {volume} {61}},\ \bibinfo {pages} {031601} (\bibinfo {year}
  {2000})}\BibitemShut {NoStop}%
\bibitem [{\citenamefont {Fan}\ \emph {et~al.}(2003)\citenamefont {Fan},
  \citenamefont {Suh},\ and\ \citenamefont {Joannopoulos}}]{fan2003temporal}%
  \BibitemOpen
  \bibfield  {author} {\bibinfo {author} {\bibfnamefont {S.}~\bibnamefont
  {Fan}}, \bibinfo {author} {\bibfnamefont {W.}~\bibnamefont {Suh}}, \ and\
  \bibinfo {author} {\bibfnamefont {J.~D.}\ \bibnamefont {Joannopoulos}},\
  }\href {\doibase 10.1364/JOSAA.20.000569} {\bibfield  {journal} {\bibinfo
  {journal} {JOSA A}\ }\textbf {\bibinfo {volume} {20}},\ \bibinfo {pages}
  {569} (\bibinfo {year} {2003})}\BibitemShut {NoStop}%
\bibitem [{\citenamefont {Maksimov}\ \emph {et~al.}(2015)\citenamefont
  {Maksimov}, \citenamefont {Sadreev}, \citenamefont {Lyapina},\ and\
  \citenamefont {Pilipchuk}}]{maksimov2015coupled}%
  \BibitemOpen
  \bibfield  {author} {\bibinfo {author} {\bibfnamefont {D.~N.}\ \bibnamefont
  {Maksimov}}, \bibinfo {author} {\bibfnamefont {A.~F.}\ \bibnamefont
  {Sadreev}}, \bibinfo {author} {\bibfnamefont {A.~A.}\ \bibnamefont
  {Lyapina}}, \ and\ \bibinfo {author} {\bibfnamefont {A.~S.}\ \bibnamefont
  {Pilipchuk}},\ }\href {\doibase 10/f7d9vm} {\bibfield  {journal} {\bibinfo
  {journal} {Wave Motion}\ }\textbf {\bibinfo {volume} {56}},\ \bibinfo {pages}
  {52} (\bibinfo {year} {2015})}\BibitemShut {NoStop}%
\bibitem [{\citenamefont {Dicke}(1954)}]{dicke1954coherence}%
  \BibitemOpen
  \bibfield  {author} {\bibinfo {author} {\bibfnamefont {R.~H.}\ \bibnamefont
  {Dicke}},\ }\href@noop {} {\bibfield  {journal} {\bibinfo  {journal} {Phys.
  Rev.}\ }\textbf {\bibinfo {volume} {93}},\ \bibinfo {pages} {99} (\bibinfo
  {year} {1954})}\BibitemShut {NoStop}%
\bibitem [{\citenamefont {Mlynek}\ \emph {et~al.}(2014)\citenamefont {Mlynek},
  \citenamefont {Abdumalikov}, \citenamefont {Eichler},\ and\ \citenamefont
  {Wallraff}}]{mlynek2014observation}%
  \BibitemOpen
  \bibfield  {author} {\bibinfo {author} {\bibfnamefont {J.~A.}\ \bibnamefont
  {Mlynek}}, \bibinfo {author} {\bibfnamefont {A.~A.}\ \bibnamefont
  {Abdumalikov}}, \bibinfo {author} {\bibfnamefont {C.}~\bibnamefont
  {Eichler}}, \ and\ \bibinfo {author} {\bibfnamefont {A.}~\bibnamefont
  {Wallraff}},\ }\href@noop {} {\bibfield  {journal} {\bibinfo  {journal} {Nat.
  Commun.}\ }\textbf {\bibinfo {volume} {5}},\ \bibinfo {pages} {1} (\bibinfo
  {year} {2014})}\BibitemShut {NoStop}%
\bibitem [{\citenamefont {Yulin}\ \emph {et~al.}(2005)\citenamefont {Yulin},
  \citenamefont {Skryabin},\ and\ \citenamefont
  {Russell}}]{yulin2005dissipative}%
  \BibitemOpen
  \bibfield  {author} {\bibinfo {author} {\bibfnamefont {A.~V.}\ \bibnamefont
  {Yulin}}, \bibinfo {author} {\bibfnamefont {D.~V.}\ \bibnamefont {Skryabin}},
  \ and\ \bibinfo {author} {\bibfnamefont {P.~S.~J.}\ \bibnamefont {Russell}},\
  }\href {\doibase 10/cndrmj} {\bibfield  {journal} {\bibinfo  {journal} {Opt.
  Express}\ }\textbf {\bibinfo {volume} {13}},\ \bibinfo {pages} {3529}
  (\bibinfo {year} {2005})}\BibitemShut {NoStop}%
\bibitem [{\citenamefont {Maksimov}\ and\ \citenamefont
  {Sadreev}(2013)}]{maksimov2013symmetry}%
  \BibitemOpen
  \bibfield  {author} {\bibinfo {author} {\bibfnamefont {D.~N.}\ \bibnamefont
  {Maksimov}}\ and\ \bibinfo {author} {\bibfnamefont {A.~F.}\ \bibnamefont
  {Sadreev}},\ }\href {\doibase 10/gf9n44} {\bibfield  {journal} {\bibinfo
  {journal} {Phys. Rev. E}\ }\textbf {\bibinfo {volume} {88}},\ \bibinfo
  {pages} {032901} (\bibinfo {year} {2013})}\BibitemShut {NoStop}%
\bibitem [{Note1()}]{Note1}%
  \BibitemOpen
  \bibinfo {note} {See Supplemental Material for derivation of coupled-modes
  equations for a system of two nonlinearly coupled RLC circuits.}\BibitemShut
  {Stop}%
\bibitem [{\citenamefont {Li}\ \emph {et~al.}(2020)\citenamefont {Li},
  \citenamefont {Wu}, \citenamefont {Huang}, \citenamefont {Lu}, \citenamefont
  {Li}, \citenamefont {Deng},\ and\ \citenamefont {Liu}}]{li2020bound}%
  \BibitemOpen
  \bibfield  {author} {\bibinfo {author} {\bibfnamefont {Z.}~\bibnamefont
  {Li}}, \bibinfo {author} {\bibfnamefont {J.}~\bibnamefont {Wu}}, \bibinfo
  {author} {\bibfnamefont {X.}~\bibnamefont {Huang}}, \bibinfo {author}
  {\bibfnamefont {J.}~\bibnamefont {Lu}}, \bibinfo {author} {\bibfnamefont
  {F.}~\bibnamefont {Li}}, \bibinfo {author} {\bibfnamefont {W.}~\bibnamefont
  {Deng}}, \ and\ \bibinfo {author} {\bibfnamefont {Z.}~\bibnamefont {Liu}},\
  }\href@noop {} {\bibfield  {journal} {\bibinfo  {journal} {Applied Physics
  Letters}\ }\textbf {\bibinfo {volume} {116}},\ \bibinfo {pages} {263501}
  (\bibinfo {year} {2020})}\BibitemShut {NoStop}%
\bibitem [{\citenamefont {Suh}\ \emph {et~al.}(2004)\citenamefont {Suh},
  \citenamefont {Wang},\ and\ \citenamefont {Fan}}]{suh2004temporal}%
  \BibitemOpen
  \bibfield  {author} {\bibinfo {author} {\bibfnamefont {W.}~\bibnamefont
  {Suh}}, \bibinfo {author} {\bibfnamefont {Z.}~\bibnamefont {Wang}}, \ and\
  \bibinfo {author} {\bibfnamefont {S.}~\bibnamefont {Fan}},\ }\href@noop {}
  {\bibfield  {journal} {\bibinfo  {journal} {IEEE J. Quantum Electron.}\
  }\textbf {\bibinfo {volume} {40}},\ \bibinfo {pages} {1511} (\bibinfo {year}
  {2004})}\BibitemShut {NoStop}%
\bibitem [{\citenamefont {Lanneb{\`e}re}\ and\ \citenamefont
  {Silveirinha}(2015)}]{lannebere2015optical}%
  \BibitemOpen
  \bibfield  {author} {\bibinfo {author} {\bibfnamefont {S.}~\bibnamefont
  {Lanneb{\`e}re}}\ and\ \bibinfo {author} {\bibfnamefont {M.~G.}\ \bibnamefont
  {Silveirinha}},\ }\href@noop {} {\bibfield  {journal} {\bibinfo  {journal}
  {Nature communications}\ }\textbf {\bibinfo {volume} {6}},\ \bibinfo {pages}
  {1} (\bibinfo {year} {2015})}\BibitemShut {NoStop}%
\bibitem [{\citenamefont {Pichugin}\ and\ \citenamefont
  {Sadreev}(2015)}]{pichugin2015frequency}%
  \BibitemOpen
  \bibfield  {author} {\bibinfo {author} {\bibfnamefont {K.~N.}\ \bibnamefont
  {Pichugin}}\ and\ \bibinfo {author} {\bibfnamefont {A.~F.}\ \bibnamefont
  {Sadreev}},\ }\href {\doibase 10/ghmjqb} {\bibfield  {journal} {\bibinfo
  {journal} {JOSA B}\ }\textbf {\bibinfo {volume} {32}},\ \bibinfo {pages}
  {1630} (\bibinfo {year} {2015})}\BibitemShut {NoStop}%
\bibitem [{\citenamefont {Bhatia}\ and\ \citenamefont
  {Szeg{\"o}}(1970)}]{bhatia1970stability}%
  \BibitemOpen
  \bibfield  {author} {\bibinfo {author} {\bibfnamefont {N.~P.}\ \bibnamefont
  {Bhatia}}\ and\ \bibinfo {author} {\bibfnamefont {G.~P.}\ \bibnamefont
  {Szeg{\"o}}},\ }\href@noop {} {{\selectlanguage {english}\emph {\bibinfo
  {title} {Stability {{Theory}} of {{Dynamical Systems}}}}}},\ Classics in
  {{Mathematics}}\ (\bibinfo  {publisher} {{Springer-Verlag}},\ \bibinfo
  {address} {{Berlin}},\ \bibinfo {year} {1970})\BibitemShut {NoStop}%
\end{thebibliography}%
 
\newpage
\onecolumngrid
\appendix
\setcounter{section}{0}
\section{Stability analysis of stationary solutions of coupled modes equations}~\label{sec:stability}
In this section, we provide details on the procedure of stability analysis of stationary solutions of Eqs.~\eqref{eq:CME_BD} and~\eqref{eq:CME_B0}. We actually use a conventional method based on the determination of eigenvalues~\cite{bhatia1970stability}. For that coupled mode equations for the bright mode and BIC can be linearized with 
\begin{equation}
  \begin{aligned}
    B = (B + m_1 e^{\lambda \tau} + n_1 e^{\lambda^* \tau}) e^{i \delta_D \tau},
    \\
    D = (D + m_2 e^{\lambda \tau} + n_2 e^{\lambda^* \tau}) e^{i \delta_D \tau},
  \end{aligned}
\end{equation}
where $m_{1,2}$ and $n_{1,2}$ are some small arbitrary functions. Therefore, the coupled-mode equations transform to
\begin{equation}
  \begin{aligned}
    \lambda m_2 = K_D m_2 + Q n_2^* + R m_1 + S n_1^*,
    \\
    \lambda^* n_2 = K_D n_2 + Q m_2^* + R n_1 + S m_1^*,
    \\
    \lambda m_1 = K_B m_1 + Q n_1^* + R m_2 + S n_2^*,
    \\
    \lambda^* n_1 = K_B n_1 + Q m_1^* + R n_2 + S m_2^*,
  \end{aligned}
\end{equation}
In combination with the conjugated versions of these equations, the matrix form can be written as
\begin{equation}
  \begin{pmatrix}
      K_B - \lambda & R & Q & S \\
      R & K_D - \lambda & S & Q \\
      Q^* & S^* & K_B^* - \lambda & R^* \\
      S^* & Q^* & R^* & K_D^* - \lambda
    \end{pmatrix}
    \begin{pmatrix}
      m_1 \\ m_2 \\ n_1^* \\ n_2^*
    \end{pmatrix}
    = 0,
\end{equation}
where the following notation is used:
\begin{equation}
  \begin{aligned}
    S &= 2i\alpha BD,
    \\
    Q &= i\alpha(B^2 + D^2),
    \\
    R &= 2i\alpha (BD^* + DB^*),
    \\
    K_B &= -i\delta_B - \Gamma_B + 2i\alpha (|B|^2 + |D|^2),
    \\
    K_D &= -i\delta_D - \Gamma_D + 2i\alpha (|B|^2 + |D|^2),
  \end{aligned}
\end{equation}
Calculation of the determinant gives the fourth-order equation for eigenvalues $\lambda$. A solution is considered to be unstable if at least one $\mathrm{Re}\lambda$  has a positive real part.

In the case of purely bright mode, when $D = 0$, the procedure is the same, but the bright mode and BIC can be considered independently. We start from the bright mode solution introducing similar linearization
\begin{equation}
  B(\tau) = (B + m e^{\lambda_B \tau} + n e^{\lambda_B^* \tau}) e^{i \delta_D \tau},
\end{equation}
leading to the following system of equations:
\begin{equation}
  \begin{pmatrix}
    K - \lambda_B & Q \\
    Q^* & K^* - \lambda_B
  \end{pmatrix}
  \begin{pmatrix}
    m \\ n^*
  \end{pmatrix}
  = 0,
  \label{eq:B0_stability_matrix}
\end{equation}
where the following notation is used:
\begin{equation}
  \begin{gathered}
    Q = i \alpha B^2,
    \\
    K = -i \delta_B - \Gamma_B + 2i\alpha |B|^2.
  \end{gathered}
\end{equation}
The above system has solutions when the matrix determinant is equal to zero. This condition allows to obtain the equation for $\lambda_B$, which is simply the quadratic equation with a solution
\begin{equation}
  \lambda_B = - \Gamma_B \pm \sqrt{- \delta_B^2 + 4 \alpha \delta_B |B|^2 - 3 \alpha^2 |B|^4}.
  \label{eq:lam_B0}
\end{equation}
The same analysis can be done for the trivial dark mode  solution. 
In this case we use
\begin{equation}
  D(\tau) = ( m e^{\lambda_D \tau} + n e^{\lambda_D^* \tau}) e^{i \delta_D \tau}.
\end{equation}
The obtained system of equations for $\lambda_D$ will be fully equivalent to the system for $\lambda_B$, hence the result differs only in parameters:
\begin{equation}
  \lambda_D = - \Gamma_D \pm \sqrt{- \delta_D^2 + 4 \alpha \delta_D |B|^2 - 3 \alpha^2 |B|^4}.
  \label{eq:lam_D0}
\end{equation}

\section{Phase portraits and frequency comb generation}
\label{sec:phase_portraits}

\begin{figure}[htbp!]
  \centering
  \includegraphics[width=\linewidth]{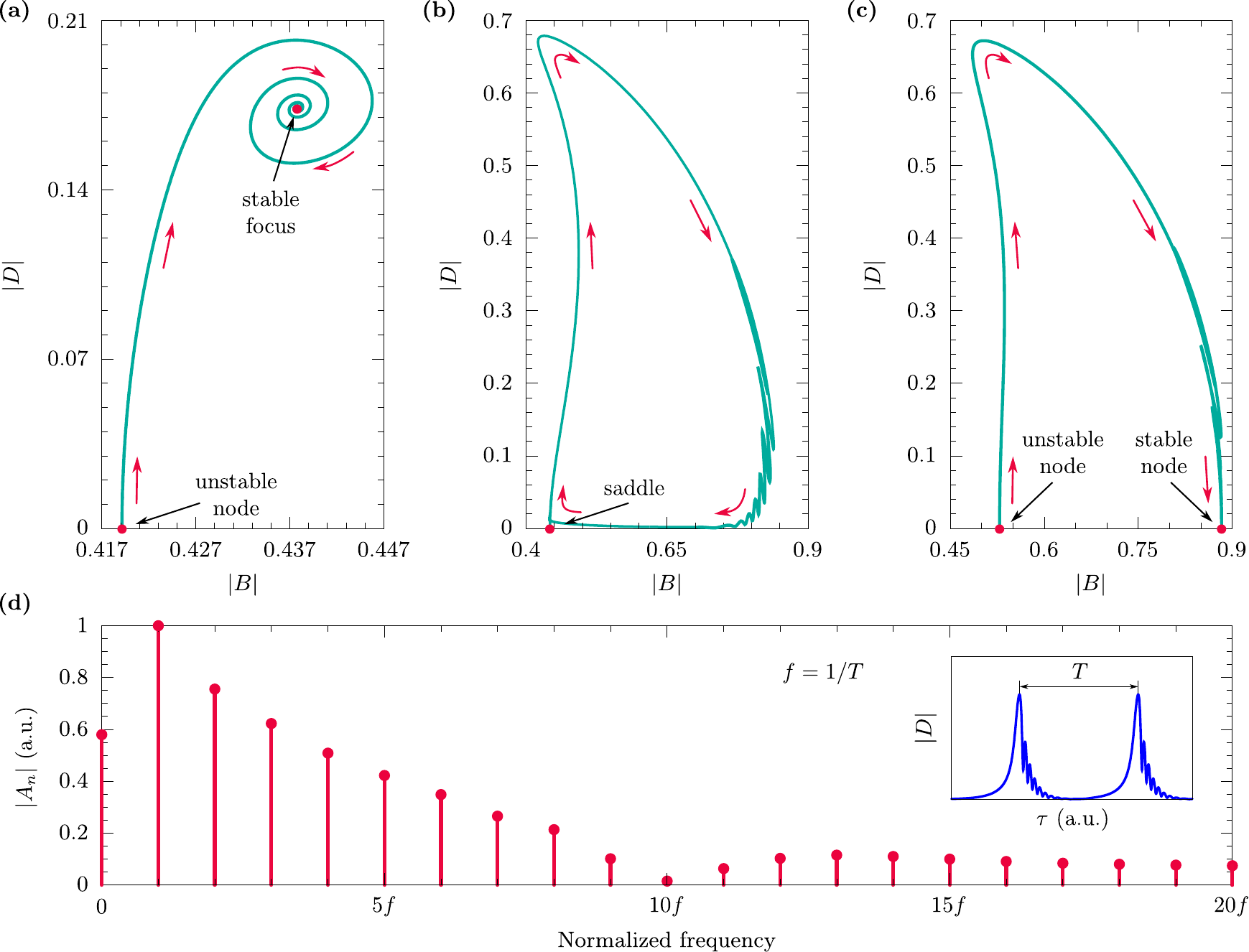}
  \caption{Phase portraits of dynamical regimes considered in Sec.~\ref{sec:dynamics}. (a) Stable focus corresponding to excitation of stable hybrid solution [Fig.~\ref{fig:D_stable}(a)], (b) closed-loop trajectory corresponding to self-oscillations [Fig.~\ref{fig:BD_self}(a)] and (c) stable node corresponding to a pulse excitation of BIC [Fig.~\ref{fig:BD_self}(b)]. System trajectories are depicted by red arrows and red dots indicates fixed points of different types. (d) Frequency comb generation: coefficients of a Fourier series for a single period of self-oscillations [Fig.~\ref{fig:BD_self}(a)]. The coefficients are normalized on a maximal value.}
  \label{fig:phase_portraits}
\end{figure}

Here, we consider phase portraits of dynamical regimes considered in Sec.~\ref{sec:dynamics}. Excitation of stable hybrid solution is associated with a stable focus point and a corresponding spiral trajectory. Example of the phase portrait for this case is demonstrated in Fig.~\ref{fig:phase_portraits}(a).
On the other hand, the self-oscillatory regime is associated with a closed-loop trajectory, as Fig.~\ref{fig:phase_portraits}(b) shows. The solution corresponding to $D=0$ is stable and therefore this point should be stable for the pure bright state. However, the presence of fluctuations in $D$ results in excitation of BIC and the system moves away from this point. Hence, trajectories in horizontal directions points to a stable point corresponding to the case when $D = 0$, but at the same time trajectories in vertical direction points out this point due to the instability of $D$ amplitude. Therefore, this point is a saddle and trajectories near it are hyperbolic. When the amplitude $D$ reaches its maximal value the effective driven force $p_D$ decreases and the system tends to return to the initial point. But in the vicinity of the saddle fluctuations in $D$ start to grow again, so the system repeats the path in phase space.
Single period of such self-oscillations is associated with generation of a large number of harmonics, as Fig.~\ref{fig:phase_portraits}(d) demonstrates, where 
\begin{equation}
    |A_n| = \sqrt{a_n^2 + b_n^2}, ~~ n = 0, 1, ...
\end{equation}
such that $a_n$ and $b_n$ are coefficients of the corresponding Fourier series:
\begin{equation}
  \begin{aligned}
    a_m &= \frac{2}{T} \int\limits_0^T F(t) \sin\left(\frac{2\pi m t}{T} \right) dt,
    &~~&
    a_0 = \frac{1}{T} \int\limits_0^T F(t) dt,
    \\
    b_m &= \frac{2}{T} \int\limits_0^T F(t) \cos \left(\frac{2\pi m t}{T} \right) dt,
    &~~&
    b_0 = 0,
  \end{aligned}
\end{equation}
where $m = 1, 2, \ldots$.
Therefore, the function in time-domain can be represented as
\begin{equation}
  F(t) = a_0 + \sum\limits_{m = 1}^{\infty} \left[a_m \sin \left(\frac{2\pi m t}{T} \right) + b_m\cos\left(\frac{2\pi m t}{T} \right) \right].
\end{equation}
Each coefficient $A_n$ is associated with the corresponding frequency $nf$, such that $f = 1/T$, where $T$ is the period of self-oscillations [see the inset of Fig.~\ref{fig:phase_portraits}(d)].
Importantly, the harmonics are equidistant and hence it can be stated that self-oscillations of the system results in the formation of a frequency comb.
Finally, single-pulse excitation of BIC corresponding to switching from one pure bright state to another pure bright state and hence the phase portrait in this case represents a movement from an unstable node to a stable one, as Fig.~\ref{fig:phase_portraits}(c) shows.

\end{document}


\title{Excitation of a bound state in the continuum via spontaneous symmetry breaking: \\
 Supplementary Materials}

\author{Alexander Chukhrov}
  \affiliation{Department of Physics and Engineering, ITMO University, Saint Petersburg 197101, Russia}%

\author{Sergey Krasikov}%
  \email{s.krasikov@metalab.ifmo.ru}
  \affiliation{Department of Physics and Engineering, ITMO University, Saint Petersburg 197101, Russia}%

\author{Alexey Yulin}
  \affiliation{Department of Physics and Engineering, ITMO University, Saint Petersburg 197101, Russia}%

\author{Andrey Bogdanov}
   \email{a.bogdanov@metalab.ifmo.ru}
  \affiliation{Department of Physics and Engineering, ITMO University, Saint Petersburg 197101, Russia}%

\date{\today}

\maketitle
\begin{figure}[htbp]
  \centering
  \includegraphics[width=\linewidth]{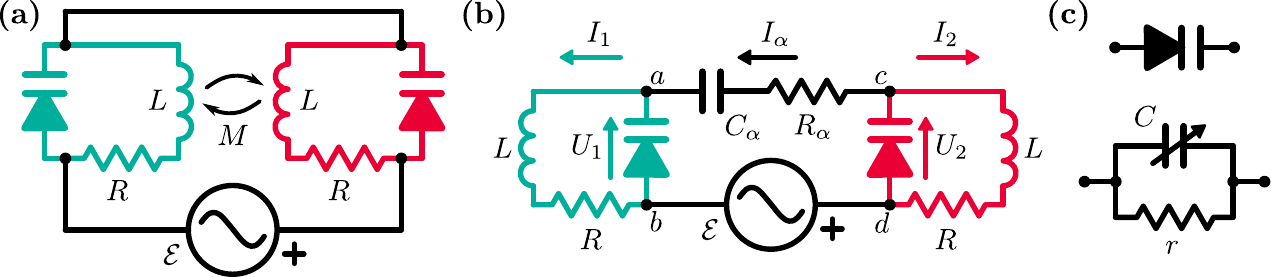}
  \caption{The system under the consideration: two coupled RLC circuits driven by the harmonic source. Coupling between the circuits may be introduce in different ways as (a) and (b) demonstrates.  (c) Equivalent scheme of the varactor diode, which consist of the capacitor $C$, series resistance $R$ and  resistance elements $r$ and $R$.}
  \label{fig:RLC}
\end{figure}
\begin{figure}[htbp!]
  \centering
  \includegraphics[width=0.5\linewidth]{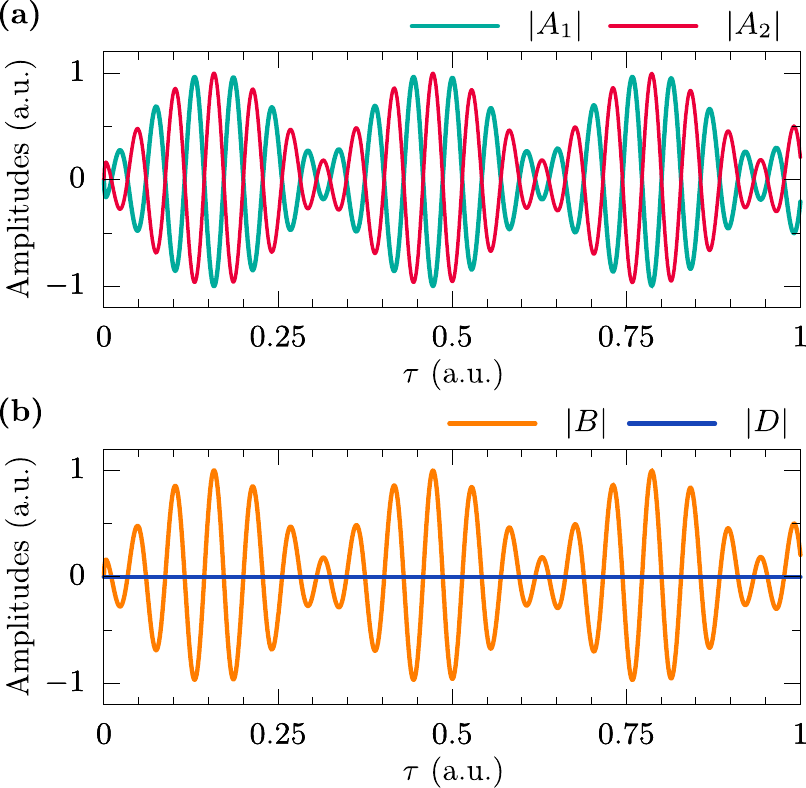}
  \caption{(a) Voltage oscillations in considered RLC circuits and (b) their interference representing the normal modes of the system.}
  \label{fig:modes}
\end{figure}

Here, we provide an example of a physical system, which may be used for the experimental verification of the obtained results. The system supporting BIC and bright node  can be implemented as two coupled RLC circuits schematically shown in Figs.~\ref{fig:RLC}(a) and~\ref{fig:RLC}(b). In the general case, the circuits are not identical but to simplify the derivation we assume them the same. Each circuit consists of the inductance $L$, resistance $R$ and varactor diodes characterized by the junction capacitance $C$ and the diode current resistance $r$. The equivalent scheme of the varactor diodes is shown in Fig.~\ref{fig:RLC}(c). The diodes are nonlinear elements such that the junction capacitance $C$ is a function of the voltage across the diode.
The circuits are driven by the voltage source with the amplitude $\mathcal{E}$. The coupling between the circuits can be accounted via a mutual inductance $M$ as shown in Fig.~\ref{fig:RLC}(a) or via a wire connection loaded with a resistance $R_{\alpha}$ and capacitance $C_{\alpha}$ as shown in Fig.~\ref{fig:RLC}(b). These ways of the coupling are equivalent, however we will consider the scheme with the wire connection [Fig.~\ref{fig:RLC}(b)] it virtue its simpler experimental realization. The differential equations describing the voltage and current distribution in this scheme can be obtained using Kirchhoff's laws:  
\begin{equation}
  \begin{aligned}
    &L \frac{d I_1}{dt} + R I_1 + U_1 = 0, &~~ I_1 &= I_{\alpha} + C \frac{d U_1}{dt} + \frac{U_1}{r}, 
    \\
    &L \frac{d I_2}{dt} + R I_2 + U_2 = 0, &~~ -I_2 &= I_{\alpha} - C \frac{d U_2}{dt} - \frac{U_2}{r},
    \\
    &\mathcal{E} = R_{\alpha} I_{\alpha} + U_{C_{\alpha}} - U_1 + U_2, &~~ I_{\alpha} &= C_{\alpha} \frac{d U_{C_{\alpha}} }{dt}.
  \end{aligned}
  \label{eq:Kirchhoff_laws_initial}
\end{equation}
Here, $I_{1,2}$ is the current through the series resistance and $U_{1,2}$ is the voltage across the varactor diodes, defined as as $U_1 = \phi_a - \phi_b$ and $U_2 = \phi_c - \phi_d$, such that $\phi_i$ ($i=a,b,c,d$) is the electric potential at the corresponding points [see Fig.~\ref{fig:RLC}(b)]. Hereinafter, we will use the dimensionless time $\tau$ normalized to the characteristic resonant frequency of the circuit $\omega_0 = 1/\sqrt{LC}$ as $\tau=\omega_0 t$.  Equations~\eqref{eq:Kirchhoff_laws_initial} can be rewritten in terms of voltages using the Ohm's law, namely $V_{1,2} = Z I_{1,2}$ and $V_{\alpha} = Z I_{\alpha}$, where $Z = \sqrt{L/C}$ is the total reactance of the circuits. The dimensionless damping coefficients related to the energy decay in resistive elements are
\begin{equation}
	\gamma_V = \frac{R}{Z}, ~~ \gamma_U = \frac{Z}{r}, ~~ \gamma_{\alpha} = \frac{C}{C_{\alpha}} \frac{Z}{R_{\alpha}},
\end{equation}
and the quality factor is $Q_{\alpha} = Z/R_{\alpha}$. In terms of these parameters, Eqs.~\eqref{eq:Kirchhoff_laws_initial} become
\begin{equation}
  \begin{aligned}
    &\dot{V}_1 = - \gamma_V V_1 - U_1, &~~& \dot{U}_1 = -\gamma_U U_1 + V_1 - V_{\alpha},
    \\
    &\dot{V}_2 = - \gamma_V V_2 - U_2, &~~& \dot{U}_2 = -\gamma_U U_2 + V_2 + V_{\alpha},
    \\
    &\dot{\tilde{U}}_{C_{\alpha}} = -\gamma_{\alpha} \tilde{U}_{C_{\alpha}} + Q_{\alpha} (U_1 - U_2) + \Pi, &~~& V_{\alpha} = \dot{\tilde{U}}_{C_{\alpha}},
  \end{aligned}
  \label{eq:Kirchhoff_laws_linear}
\end{equation}
where $\tilde{U}_{C_{\alpha}} = (C_{\alpha} /C) U_{C_{\alpha}}$ and $\Pi = Q_{\alpha} \mathcal{E}$. The dot means time derivative over the dimensionless time $\tau$. 
The obtained equations allow to describe the linear system, but we have not accounted for nonlinearity occurring due to the presence of varactor diodes. For that the charge of the junction capacitor and hence the corresponding current can be expanded in a series as
\begin{equation}
  \begin{gathered}
    q_{1,2} \approx C U_{1,2} + \eta_1 C U_{1,2}^2 + \eta_2 C U_{1,2}^3, 
    \\
    I_{1,2} = \dot{q}_{1,2} \approx C (1 + \eta_1 U_{1,2} + \eta_2 U_{1,2}^2) \dot{U}_{1,2}.
  \label{eq:varicap_charge}
  \end{gathered}
\end{equation}
To incorporate this expansion into the Eqs.~\eqref{eq:Kirchhoff_laws_linear}, we introduce $\chi_1 = - \eta_1$ and $\chi_2 = \eta_1^2 - \eta_2$, which comes from the following expansion
\begin{equation}
  \frac{1}{1 + \eta_1 U + \eta_2 U^2} \approx 1 + \chi_1 U + \chi_2 U^2.
\end{equation}
Finally, the equations can be written as
\begin{equation}
  \begin{aligned}
    &\dot{V}_1 + U_1= - \gamma_V V_1, &~~& \dot{U}_1 - V_1 = -\gamma_U U_1 - V_{\alpha} + V_1 (\chi_1 U_1 + \chi_2 U_1^2),
    \\
    &\dot{V}_2 + U_2 = - \gamma_V V_2, &~~& \dot{U}_2 - \tilde{V}_2 = -\gamma_U U_2 + V_{\alpha} + V_2 (\chi_1 U_1 + \chi_2 U_1^2),
    \\
    &\dot{\tilde{U}}_{C_{\alpha}} = -\gamma_{\alpha} \tilde{U}_{C_{\alpha}} + Q_{\alpha} (U_1 - U_2) + \Pi, &~~& V_{\alpha} = \dot{\tilde{U}}_{C_{\alpha}}.
  \end{aligned}
  \label{eq:Kirchhoff_laws_nonlinear}
\end{equation}
Assuming that all terms on the right-hand sides are small, we can solve them using perturbational approach, deriving coupled-mode equations within the slow varying amplitudes approximation. For that the equations can be written as 
\begin{equation}
	\dot{\mathbf{Y}} = (\hat{H}_0 + \hat{H}_p) \mathbf{X},
\end{equation}
where $\mathbf{Y} = A(\tau) \mathbf{X(\tau)}$. Therefore, the system can be separated into two parts, perturbed and unperturbed:
\begin{equation}
	\begin{aligned}
		\dot{A}(\tau) \mathbf{X} = \hat{H}_p \mathbf{X}, 
		\\
		A(\tau) \dot{\mathbf{X}} = \hat{H}_0 \mathbf{X}, 
	\end{aligned}
	\label{eq:system_separation}
\end{equation}
such that $\mathbf{X}$ is the solution of the unperturbed part, which for the considered case can be written as
\begin{equation}
	\begin{aligned}
		\dot{U}_i - V_i = 0
		\\
		\dot{V}_i + U_i = 0
	\end{aligned}
	~~
	\Rightarrow
	~~
	\dot{\mathbf{U}}_i
	=
	\begin{pmatrix}
		0 & 1 \\
		-1 & 0
	\end{pmatrix}
	\mathbf{U}_i, 
	~~
	\mathbf{U}_i = 
	\begin{pmatrix}
		U_i \\ V_i
	\end{pmatrix},
	\label{eq:unperturbed system}
\end{equation}
where $i = 1,2$. The solution of this system is simply
\begin{equation}
	\mathbf{U}_i = A_i \boldsymbol{\xi} e^{i\tau} + A_i^* \boldsymbol{\xi}^* e^{-i\tau},
	\label{eq:solution_unperturbed}
\end{equation}
such that $\boldsymbol{\xi} = (1, i)^T$ and $^*$ stands for complex conjugate. However, 
we use the \textit{slowly varying amplitude approximation}, implying that only $A_i \boldsymbol{\xi} e^{i\tau}$ has to be considered. So the solution of the unperturbed system is
\begin{equation}
	\mathbf{U}_i = A_i \boldsymbol{\xi} e^{i\tau}.
\end{equation}
Then the perturbations have to be taken into account. According to Eq.~\eqref{eq:system_separation}, the contribution from the perturbations can be calculated as
\begin{equation}
	\dot{A}_i \boldsymbol{\xi} e^{i\tau} = \hat{H}_p \boldsymbol{\xi} e^{i\tau}.
\end{equation}
Both sides of this expression can be multiplied by $\boldsymbol{\xi}^{\dagger} = (1, -i)$ from the left, so the corrections for the amplitude can be written in the explicit form:
\begin{equation}
	\dot{A}_i = \frac{1}{2} \boldsymbol{\xi}^{\dagger} \hat{H}_p \boldsymbol{\xi}.
\end{equation}
The operator $\hat{H}_p$ can be separated into two parts corresponding to the linear and nonlinear perturbations, which can be considered independently due to the linearity of the equations. Equations~\eqref{eq:Kirchhoff_laws_nonlinear} written for linear perturbations  are
\begin{equation}
  \begin{aligned}
    &\dot{V}_1 + U_1= - \gamma_V V_1, &~~& \dot{U}_1 - V_1 = -\gamma_U U_1 - V_{\alpha},
    \\
    &\dot{V}_2 + U_2 = - \gamma_V V_2, &~~& \dot{U}_2 - V_2 = -\gamma_U U_2 + V_{\alpha} .
  \end{aligned}
\end{equation}
Here, we also can consider the contributions from $\gamma_V$, $\gamma_U$ and $V_{\alpha}$ independently. Hence, we can start from the system
\begin{equation}
  	\begin{pmatrix}
  		\dot{U}_i - V_i 
  		\\
  		\dot{V}_i + U_i
  	\end{pmatrix}
  	=
  	\begin{pmatrix}
  		-\gamma_U & 0 \\
  		0 & -\gamma_V
  	\end{pmatrix}
  	\begin{pmatrix}
  		U_i 
  		\\
  		V_i
  	\end{pmatrix},
\end{equation}
from where we can easily obtain
\begin{equation}
	\frac{1}{2} \boldsymbol{\xi}^{\dagger} \hat{H}_p \boldsymbol{\xi} = - \frac{\gamma_U + \gamma_V}{2}.
\end{equation}
To consider the correction related to $V_{\alpha}$, we firstly need to found $\dot{\tilde{U}}_{C_{\alpha}}$, since $V_{\alpha} = \dot{\tilde{U}}_{C_{\alpha}}$ according to Eqs.~\eqref{eq:Kirchhoff_laws_nonlinear}, from which we can write as
\begin{equation}
	\tilde{U}_{C_{\alpha}} = \frac{1}{\gamma_{\alpha} + i} (Q_{\alpha}(A_1 - A_2) + g) e^{i \tau} + U_0 e^{-\gamma_{\alpha} \tau},
\end{equation}
where $\Pi = g e^{i \tau}$ and $U_0$ is the constant arising from the solution of the corresponding equation for $\dot{\tilde{U}}_{C_{\alpha}}$. The term $U_0 e^{-\gamma_{\alpha} \tau}$ tends to zero at large $\tau$ and, therefore, it can be neglected. Thus, we can find the next correction to the amplitude:
\begin{equation}
	\frac{1}{2} \boldsymbol{\xi}^{\dagger} 
	\begin{pmatrix}
		V_{\alpha}
		\\
		0
	\end{pmatrix}
	=
	\frac{1}{2} \frac{i}{\gamma_{\alpha} + i} (Q_{\alpha}(A_1 - A_2) + g).
\end{equation}
At this stage, the coupled mode equations accounting for linear perturbations can be written  as
\begin{equation}
	\begin{aligned}
		\dot{A}_1 = -\Gamma_0 A_1 - \frac{i Q_{\alpha}}{\gamma_{\alpha} + i} \frac{A_1 - A_2}{2} - p e^{i \delta_0 \tau},
		\\
		\dot{A}_2 = -\Gamma_0 A_2 + \frac{i Q_{\alpha}}{\gamma_{\alpha} + i}  \frac{A_1 - A_2}{2} + p e^{i \delta_0 \tau},
	\end{aligned}
\end{equation}
where the decay rate $\Gamma_0$ and the pump amplitude $p$ are 
\begin{equation}
	\Gamma_0 = \frac{\gamma_U + \gamma_V}{2} ~~\text{and} ~~ p = \frac{i}{\gamma_{\alpha} + i} \frac{g_0}{2}.
\end{equation}
Here $g = g_0 e^{i \delta_0 \tau}$, and $\delta_0$ is the detuning of the pump frequency $\omega$ from the characteristic resonant frequency of the circuit $\omega_0$. These equations still need to be supplemented by terms corresponding to nonlinear perturbations. Namely, we need to consider the system
\begin{equation}
	\begin{aligned}
		\dot{U}_i - V_i &= (\chi_1 U_i + \chi_2 U_i^2) V_i,
		\\
		\dot{V}_i + U_i &= 0,
	\end{aligned}
	\label{eq:nonlin_pert_system}
\end{equation}
The corrections related to $\chi_1 U_i V_i$ and $\chi_2 U_i^2 V_i$ can be considered independently. To find them we will use perturbation approach
\begin{equation}
	\begin{aligned}
		U_i \approx U_i' + U_i'', 
		\\
		V_i \approx V_i' + V_i''.
	\end{aligned}
\end{equation} 
Here, $'$ and $''$ indicate the order of smallness, such that $U_i'' \ll U_i'$ and $V_i'' \ll V_i'$. Therefore, to find the corrections related to $\chi_1 U_i V_i$ term, the system of Eqs.~\eqref{eq:nonlin_pert_system} transforms to
\begin{equation}
	\begin{aligned}
		\dot{U}_i - V_i &= \chi_1 U_i V_i,
		\\
		\dot{V}_i + U_i &= 0,
	\end{aligned}
	~~
	\Rightarrow
	~~
	\begin{aligned}
		&\dot{U}_i' - V_i' + \dot{U}_i'' - V_i''  = \chi_1 U_i' V_i'  ,
		\\
		&\dot{V}_i' + U_i' + \dot{V}_i''  + U_i'' = 0.
	\end{aligned}
\end{equation}
The obtained equations can be separated into two subsystems corresponding to different orders of smallness:
\begin{equation}
		\begin{aligned}
		&\dot{U}_i' - V_i' = 0 ,
		\\
		&\dot{V}_i' + U_i' = 0,
	\end{aligned}
	~~~
	\text{and}
	~~~
	\begin{aligned}
		& \dot{U}_i'' - V_i''  = \chi_1 U_i' V_i'  ,
		\\
		& \dot{V}_i''  + U_i'' = 0,
	\end{aligned}
\end{equation}
However, the solutions for the first order subsystem are known, since $U_i' = Ae^{i\tau} + A^*e^{-i\tau}$ is just a solution of the unperturbed system~\eqref{eq:unperturbed system}. Therefore, using the equation $\dot{U}_i' - V_i' =0$, we can find $V_i' = iAe^{i\tau} - iA^*e^{-i\tau}$. We also can find the solution for the second order perturbations:
\begin{equation}
	\begin{aligned}
		U_i'' = \frac{2}{3} \chi_1 \left(A^2 e^{2i\tau} + (A^*)^2 e^{-2i\tau} \right),
		\\
		V_i'' = \frac{i}{3} \chi_1 \left(A^2 e^{2i\tau} - (A^*)^2 e^{-2i\tau} \right).
	\end{aligned}
\end{equation}
Here we also can use the slowly varying amplitude approximation assuming that only $\propto e^{i\tau}$ terms have to be considered. This means that 
\begin{equation}
	 (\chi_1 U_i + \chi_2 U_i^2) V_i \approx \chi_1 (U_i' V_i'' + U_i'' V_i') + \chi_2 (U_i')^2 V_i' = -\frac{i}{3} \chi_1^2 A_i |A_i|^2 + i\chi_2 A_i |A_i|^2,
\end{equation}
since all other terms can be cancelled out. The contribution from the nonlinear perturbations then can be found as
\begin{equation}
	\boldsymbol{\xi}^{\dagger} =
	\begin{pmatrix}
		\chi_1 UV + \chi_2 U^2 V 
		\\
		0
	\end{pmatrix}
	\approx
	i \left(\chi_2 - \frac{\chi_1^2}{3} \right) A_i |A_i|^2,
\end{equation}
so the final coupled-mode equations are
\begin{equation}
	\begin{aligned}
		\dot{A}_1 = -\Gamma_0 A_1 - \frac{i Q_{\alpha}}{\gamma_{\alpha} + i} \frac{A_1 - A_2}{2} + i \alpha A_1 |A_1|^2 - p e^{i \delta_0 \tau},
		\\
		\dot{A}_2 = -\Gamma_0 A_2 + \frac{i Q_{\alpha}}{\gamma_{\alpha} + i}  \frac{A_1 - A_2}{2} + i \alpha A_2 |A_2|^2 + p e^{i \delta_0 \tau}.
		\label{eq:CME_A12}
	\end{aligned}
\end{equation}
Here $\alpha =  \chi_2 - \chi_1^2/3$ is the nonlinear coefficient. The solutions of these equations are shown in Fig.~\ref{fig:modes}(a).
However, the voltage oscillations in circuits actually do not represent normal modes of the system which are formed by interference of these oscillations. Therefore, amplitudes of the normal modes can be presented as
\begin{equation}
  B = \frac{A_2 - A_1}{2}, ~~ D = \frac{A_2 + A_1}{2}.
  \label{eq:BD_amplitudes}
\end{equation}
Since $A_1$ and $A_2$ differ only by phase, $B$ and $D$ correspond to symmetric and anti-symmetric modes, or bright mode and BIC, as it was discussed in the main text. Figure~\ref{fig:modes}(b) explicitly demonstrates that the BIC is not excited by the external pump. It is important to note that the graphs in Fig.~\ref{fig:modes} are plotted for the case when there is no damping in the system and beating occur due to the difference between the frequency of the pump $\omega$ and the frequency of the bright mode $\omega_B$.
In the basis of amplitudes of bright mode and BIC, Eqs.~\eqref{eq:CME_A12} become
\begin{equation}
  \begin{aligned}
    \dot{D} &= - \Gamma_0 D + i \alpha \left[ D \left(|D|^2 + 2 |B|^2\right) + B^2 D^* \right],
  \\
    \dot{B} &= - \Gamma_0 B - \frac{i Q_{\alpha}}{\gamma_{\alpha} + i} B + i \alpha \left[ B \left(|B|^2 + 2 |D|^2\right) + D^2 B^* \right] + p e^{i \delta_0 \tau}.
  \end{aligned}
\end{equation}
Then, without the loss of generality, the complex amplitudes can be presented as
\begin{equation}
  B \rightarrow B(\tau) e^{i \delta_0 \tau}, ~~ D \rightarrow D(\tau) e^{i\delta_0 \tau}.
\end{equation}
Thus, the above equations transform to
\begin{equation}
  \begin{aligned}
    \dot{D} &= - i \delta_D D - \Gamma_D D + i \alpha D \left(|D|^2 + 2 |B|^2\right) + i \alpha B^2 D^*,
    \\
    \dot{B} &= - i \delta_B B - \Gamma_B B + i \alpha B \left(|B|^2 + 2 |D|^2\right) + i \alpha D^2 B^* + p,
  \end{aligned}
\end{equation}
where the following parameters were introduced:
\begin{equation}
  \begin{aligned}
    \delta_B &= \delta_0 + \frac{\gamma_{\alpha} Q_{\alpha}}{\gamma_{\alpha}^2 + 1}, & ~~~ &\delta_D = \delta_0,
    \\
    \Gamma_B &= \Gamma_0 + \frac{Q_{\alpha}}{\gamma_{\alpha}^2 + 1}, & ~~~ &\Gamma_D = \Gamma_0.
  \end{aligned}
    \label{eq:BD_params}
\end{equation}
Hence, we have obtained the Eqs.~(1) which are used in the main text. This suggests that the considered RLC circuits may be used as a platform for the experimental verification of obtained results.